\documentclass[floatfix,prb,showpacs,reprint]{revtex4-1}
\usepackage{graphicx}
\usepackage{amsfonts}
\usepackage{amssymb}
\usepackage{amsmath}
\usepackage{bm}% bold math
\usepackage{bbold}
\usepackage{color}
\usepackage{physics}
\usepackage{nicefrac}
\usepackage{xr}
\usepackage{tikz}
\usetikzlibrary{shapes,arrows.meta,shapes.geometric}
\newcommand{\be}{\begin{equation}}
\newcommand{\ee}{\end{equation}}
\newcommand{\bea}{\begin{eqnarray}}
\newcommand{\eea}{\end{eqnarray}}
\newcommand{\Eq}[1]{Eq.\,(\ref{#1})}% \Eq{abc}
\newcommand{\Fig}[1]{Fig.\,\ref{#1}}% \Fig{fig:abc}

\newcommand{\App}[1]{Appendix\,\ref{#1}}
\newcommand{\Vol}{\mathbb{V}}%

\begin{document}

\title{Phonon-Induced Dephasing in Quantum Dot-Cavity QED}

\author{A. Morreau}
\email[Electronic address: ]{morreauai@cardiff.ac.uk}
\author{E. A. Muljarov}
\affiliation{%
School of Physics and Astronomy, Cardiff University, Cardiff CF24 3AA, United Kingdom}
%\pacs{,,}
\date{\today}

\begin{abstract}

We present a semi-analytic and asymptotically exact solution to the problem of phonon-induced decoherence in a quantum dot-microcavity system. Particular emphasis is placed on the linear polarization and optical absorption, but the approach presented herein may be straightforwardly adapted to address any elements of the exciton-cavity density matrix. 
At its core, the approach combines Trotter's decomposition theorem with the linked cluster expansion. The effects of the exciton-cavity and exciton-phonon couplings are taken into account on equal footing, thereby providing access to regimes of comparable polaron and polariton timescales. We show that the optical decoherence is realized by real phonon-assisted transitions between different polariton states of the quantum dot-cavity system, and that the polariton line broadening is well-described by Fermi's golden rule in the polariton frame. We also provide purely analytic approximations which accurately describe the system dynamics in the limit of longer polariton timescales.
\end{abstract}

\maketitle

A quantum dot (QD) embedded in a solid-state optical microcavity presents a fundamental system within cavity quantum electrodynamics (cavity-QED)~\cite{HennessyNat07}. The QD exciton couples to an optical mode of the cavity in a manner well described by the exactly solvable Jaynes-Cummings (JC) model~\cite{JaynesIEEE63,VallePRB09,KasprzakNatMat10}. Within the strong coupling regime there is a partly reversible exchange of energy, with a period $\tau_{\rm JC}$, between the exciton and the cavity mode, which gives rise to {\em polariton} formation and characteristic vacuum Rabi splitting~\cite{ThompsonPRL92,ReitzensteinJPhysD10,OtaAPL18}.

Whilst not accounted for in the JC model, there is significant experimental and theoretical evidence~\cite{Wilson-RaePRB02,McCutcheon10,Ota09,KaerPRL10,HohenesterPRB10,RoyPRL11,GlasslPRB12,NazirJPCM16,NahriJPCM16,HorneckerPRB17,HohenesterPRB09,CalicPRL11,ValentePRB14,PortalupiNL15,MullerPRX15} to suggest that phonons play a crucial role in the optical decoherence of the QD-cavity system. The general phenomenon of phonon-induced dephasing in semiconductor QDs is well studied; it has been successfully explained and quantified by the exactly solvable independent boson (IB) model~\cite{Mahan00}. This model describes a {\em polaron}, formed from a QD exciton coupled to bulk acoustic phonons~\cite{KrummheuerPRB02}, with a  characteristic polaron formation time $\tau_{\rm IB}$.  The IB model accounts for the major effect of the non-Markovian pure dephasing but is known to fail treating the exciton zero-phonon line (ZPL) broadening~\cite{MuljarovPRL04}.

It is natural to draw upon the JC and IB models when addressing the problem of phonon-induced dephasing in the QD-cavity system. However, the combination of the two models presents a significant challenge. Various approaches to the QD-cavity problem have been suggested in the literature, ranging from Born-Markov approximations~\cite{Wilson-RaePRB02,KaerPRL10,McCutcheon10} to path-integral methods~\cite{GlasslPRB12,NahriJPCM16,MakarovCPL94,SimJCP01,GlasslPRB12,VagovPRL07,VagovPRL14,Cygorek2017} and non-equilibrium Green's function techniques~\cite{HorneckerPRB17}. These approaches can be broadly divided into perturbative and non-perturbative methods.

The perturbative methods employ a polaron transformation followed by a perturbative treatment of the coupling of the phonon-dressed exciton to the cavity mode, carried out in the
2nd order Born approximation~\cite{Wilson-RaePRB02,KaerPRL10,Ota09,McCutcheon10} or beyond~\cite{NazirJPCM16, HorneckerPRB17}. These approaches perform well in certain parameter regimes but break down, for example, when the polaron formation time $\tau_{\rm IB}$ is comparable to, or slower than, the exciton-cavity oscillation period of the polariton $\tau_{\rm JC}$.

Non-perturbative techniques based on a quasi-adiabatic Feynman path-integral scheme~\cite{MakarovCPL94} enable accurate numerical solutions but are computationally expensive and provide little insight into the underlying physics.  Nahri \textit{et al.}~\cite{NahriJPCM16} apply a tensor multiplication scheme~\cite{MakarovCPL94} to the case of a QD-cavity system with super-ohmic spectral density. This technique relies upon a complex algorithm with an ``on-the-fly path selection'' optimization~\cite{SimJCP01}. Glassl \textit{et al.}~\cite{GlasslPRB12} present a real-time path-integral scheme~\cite{VagovPRL07} adapted for a QD in a lossless cavity. Cavity and QD dampings are included within later work~\cite{VagovPRL14}, but in this case the exciton-phonon coupling is added phenomenologically.

In this paper, we present a semi-analytic exact solution of the long-standing problem of the phonon-induced decoherence of the QD-cavity system. Our approach is based on the Trotter decomposition with a subsequent use of the cumulant expansion technique~\cite{Mahan00, MuljarovPRL04, MuljarovPRL05}, which provides a computationally straightforward and physically intuitive formulation. Being non-perturbative, our approach treats the effects of the exciton-photon and exciton-phonon couplings on equal footing, thereby rendering the technique appropriate across the full range of both coupling strengths, as well as timescales $\tau_{\rm IB}$ and $\tau_{\rm JC}$. We additionally provide a physical interpretation of our findings based on a theoretically rigorous polariton model.
%Furthermore, the solutions we present become fully analytical in the limit $\tau_{\text{IB}}\ll\tau_{\text{JC}}$.

A key principle of the present method is a separation of the system Hamiltonian into two exactly solvable parts, $H = H_{\rm JC} + H_{\rm IB}$, described by the JC and IB models respectively. The JC Hamiltonian has the form ($\hbar=1$):
\begin{equation}
H_{\rm JC} = \omega_X d^\dagger d + \omega_C a^\dagger a + g(a^\dagger d + d^\dagger a)\,,
\label{eq:HJC}
\end{equation}
where $d^{\dagger}$ ($a^{\dagger}$) is the exciton (cavity photon) creation operator, $g$ is the exciton-cavity coupling strength, and $\omega_X$ ($\omega_C$) is the exciton (cavity photon) complex frequency,
\begin{equation}
\omega_{X,C} = \Omega_{X,C} - i\gamma_{X,C}\,.
\end{equation}
The imaginary frequency component $\gamma_X$ ($\gamma_C$) characterizes the long-time ZPL exciton dephasing (cavity mode radiative decay) rate. Note that this non-Hermitian Hamiltonian $H_{\rm JC}$ is straightforwardly derived from its Hermitian analog through the Lindblad dissipator formalism, as shown in \App{sec:compHamiltonian}. 

For convenience, the ZPL term from the standard IB Hamiltonian~\cite{Mahan00} can been included within $H_{\rm JC}$, \Eq{eq:HJC}, giving $H_{\rm IB}$ of the form:
\begin{equation}
H_{\rm IB} =  H_{\rm ph} + d^\dagger d V\,,
\end{equation}
where $H_{\rm ph}$ is the free phonon bath Hamiltonian and $V$ describes the exciton-phonon interaction,
\begin{equation}
H_{\rm ph} = \sum_{q}\omega_q b_q^\dagger b_q\,,\ \ \ \
V = \sum_q \lambda_q (b_q + b_{-q}^\dagger)\,.
\label{eq:HIB}
\end{equation}
Here, $b_q^\dagger$ ($\omega_q$) is the creation operator (frequency) of the $q$-th phonon mode and $\lambda_q$ is the matrix element of the exciton-phonon coupling.

It is instructive, at this point, to formally introduce timescales $\tau_{\rm JC}$ and $\tau_{\rm IB}$ associated with the JC and IB Hamiltonians respectively. The polariton timescale $\tau_{\rm JC}$ characterizes the temporal period of the Rabi oscillations,
\begin{equation}
\tau_{\rm JC} = \frac{2 \pi}{\Delta \omega}\,,\label{eq:tauJC}
\end{equation}
where $\Delta \omega$ is the polariton line separation. In the absence of phonons and for the case of zero detuning, $\Omega_X = \Omega_C$, the polariton Rabi splitting is simply twice the exciton-cavity coupling strength: $\Delta \omega = 2g$.

We define the polaron timescale as
\be
\tau_{\rm IB}\approx \sqrt{2}\pi l/v_s\,,
\label{eq:tauIB}
\ee
where $l$ is the exciton confinement radius and $v_s$ is the sound velocity, Throughout this work, we take $l=3.3$ nm and $v_s = 4.6\,\times 10^3$ m/s. Note that \Eq{eq:tauIB} underestimates the polaron timescale at very low temperatures ($\lesssim 5$ K) - see \App{Sec:IBLong} for further discussion. Physically, the polaron timescale characterizes the time to form (disperse) a polaron cloud following creation (destruction) of an exciton.

Whilst our approach is general and suited for describing the dynamics of any elements of the reduced density matrix of the JC sub-system, in this paper we concentrate on the most simple and intuitively clear quantity: the linear optical polarization. For this purpose, it is sufficient to reduce the basis of the JC system to the following three states: the absolute ground state $\ket{0}$, the excitonic excitation $\ket{X}$, and the cavity excitation $\ket{C}$. In this basis, $d^{\dagger} = \ket{X}\bra{0}$ and $a^{\dagger} = \ket{C}\bra{0}$.
The linear polarization is then given by a $2\!\times \!2$ matrix $\hat{P}(t)$ with the matrix elements $P_{jk}(t)$ expressed in terms of the time evolution operator $\hat{U}(t)$ as
\begin{equation}
P_{jk}(t) = \langle \bra{j}\hat{U}(t)\ket{k}\rangle_{\rm ph}\,, \ \ \ \ \
\hat{U}(t) = e^{iH_{\rm ph}t}e^{-iHt}\,,
\label{eq:Pdef}
\end{equation}
where $\langle \dots\rangle_{\rm ph}$ denotes the expectation value over all phonon degrees of freedom in thermal equilibrium and $j,k = X,C$, see \App{sec:compHamiltonian} for details. Here, $j$ indicates the initial excitation mode of the system and $k$ the mode in which the polarization is measured. For example, $P_{XX}$ ($P_{CC}$) denotes the excitonic (photonic) polarization under a pulsed exciton (cavity) excitation.

Using Trotter's decomposition theorem, the time evolution operator $\hat{U}(t)$ can be re-expressed as
\begin{align}
\hat{U}(t) &= \lim_{\Delta t \to 0} e^{iH_{ph}t} \left(e^{-i H_{\text{IB}} \Delta t}e^{-i H_{\text{JC}} \Delta t}\right)^N\,, \label{eq:Trotter}
\end{align}
where $\Delta t = t/N$. We introduce two new operators, $\hat{M}$ and $\hat{W}$, associated with the JC and IB Hamiltonians, respectively,
\begin{align}
\hat{M}(t_n - t_{n-1}) &= \hat{M}(\Delta t) = e^{-i H_{\text{JC}} \Delta t} \label{eq:M},\\
\hat{W}(t_n,t_{n-1}) &= e^{i H_{\rm ph} t_n}e^{-i H_{\text{IB}} \Delta t}e^{-i H_{\rm ph} t_{n-1}},
\end{align}
where $t_n = n\Delta t$. Exploiting the commutivity of $H_{\text{JC}}$ and $H_{\rm ph}$ enables us to express the time evolution operator as
a time-ordered product of pairs $\hat{W}\hat{M}$:
\begin{equation}
\hat{U}(t) = \mathcal{T} \prod_{n=1}^{N} \hat{W}(t_n,t_{n-1})\hat{M}(t_n - t_{n-1}), \label{eq:Ut}
\end{equation}
where $\mathcal{T}$ is the time ordering operator. Noting that both $\hat{W}$ and $\hat{M}$ are $2\!\times \!2$ matrices in the $\ket{X}$, $\ket{C}$ basis and that $\hat{W}$ is diagonal (with diagonal elements $W_i$), the polarization \Eq{eq:Pdef} takes the form
\begin{align}
P_{jk}(t) &= \sum_{i_{N-1} = X,C} \cdots \sum_{i_1=X,C} M_{i_N i_{N-1}} \cdots M_{i_2 i_1} M_{i_1 i_0} \nonumber \\
& \times \left\langle W_{i_N}(t,t_{N-1}) \cdots W_{i_2}(t_2,t_1)W_{i_1}(t_1,0)\right\rangle_{\rm ph}\,, \label{eq:Pjk}
\end{align}
where $i_N =j$, $i_0= k$, $M_{i_n i_m} = [\hat{M}(\Delta t)]_{i_n i_m}$, and
\begin{equation}
{W}_{i_n}(t_n,t_{n-1}) =
\mathcal{T} \exp{-i \delta_{i_n X} \int_{t_{n-1}}^{t_n} V(\tau) d\tau}
\end{equation}
with $\delta_{i j}$ the Kronecker delta and $V(\tau) = e^{iH_{\rm ph}\tau}Ve^{-iH_{\rm ph}\tau}$. Further details and intermediate steps are provided in \App{sec:trotterU}.

It is instructive at this point to introduce the concept of a ``realization" of the system as a particular combination of indices $i_n$ within the full summation of \Eq{eq:Pjk}. We associate with each realization a step-function $\hat{\theta}(\tau)$ being equal to 0 over the time interval $t_n - t_{n-1}$ if $i_n = C$ (the system is in the cavity state $\ket{C}$) or 1 if $i_n = X$ (the system is in the excitonic state $\ket{X}$).  An example realization is given in \App{sec:NN_example}. The product of $W$-operators for a particular realization can be written as
\begin{equation}
W_{i_N}(t,t_{N-1}) \cdots W_{i_1}(t_1,0) = \mathcal{T} \exp{-i\int_{0}^{t} \bar{V}(\tau) d\tau}\,, \label{eq:productW}
\end{equation}
where $\bar{V}(\tau) = \hat{\theta}(\tau)V(\tau)$. Now, applying the linked cluster theorem~\cite{Mahan00} for calculating the trace of \Eq{eq:productW} over all phonon states,
we obtain
\begin{equation}
\left \langle W_{i_N}(t,t_{N-1}) \cdots W_{i_2}(t_2,t_1)W_{i_1}(t_1,0) \right\rangle_{\rm ph} = e^{\bar{K}(t)}, \label{eq:expWlincum}
\end{equation}
where
\begin{equation}
\bar{K}(t) = - \frac{1}{2} \int_0^t d\tau_1 \int_0^t d\tau_2 \langle \mathcal{T} \bar{V}(\tau_1) \bar{V}(\tau_2)\rangle
\label{eq:lincum}
\end{equation}
is the linear cumulant for the particular realization. Its explicit dependence on the specific indices $i_n$ of the realization is
given by
\begin{equation}
\bar{K}(t) = \sum_{n=1}^N \sum_{m=1}^N \delta_{i_n X}\delta_{i_m X} K_{|n - m|}\,, \label{eq:Ksum}
\end{equation}
where
\begin{equation}
K_{|n - m|} = - \frac{1}{2} \int_{t_{n-1}}^{t_n} d\tau_1 \int_{t_{m-1}}^{t_m} d\tau_2 \langle \mathcal{T} V(\tau_1)V(\tau_2)\rangle.\label{eq:Knm}
\end{equation}
Note that $K_{|n - m|}$ depends only on the time difference $|t_n - t_m| = \Delta t |n-m|$. Furthermore, as shown in \App{Sec:Cumulant}, all $K_{|n - m|}$ can be efficiently calculated from the standard IB model cumulant \mbox{$K(t) = \mathcal{T} \exp{-i\int_0^t V(\tau) d\tau}$} (calculation of the latter is detailed in Appendices \ref{Sec:IBLong} and \ref{Sec:Matrix_elements}).

Having in mind an application of this theory to semiconductor QDs coupled to bulk acoustic phonons, we use the conditions of the super-Ohmic coupling spectral density and a finite phonon memory time~\cite{VagovPRL07}. This permits a dramatic reduction in the number of terms within the double summation of \Eq{eq:Ksum}. Indeed, we need to take into account only instances in which $|t_m - t_n| \leqslant \tau_{\text{IB}}$. When selecting $\Delta t$, we must also be mindful of the requirement imposed by the Trotter decomposition method: $\Delta t \rightarrow 0$. In practice, $\Delta t$ must simply be small relative to the period of oscillation between exciton and cavity states $\tau_{\rm JC}$.

We initially consider the most straightforward application of the technique, which will be referred to as the nearest neighbors (NN) approach.

In the NN approach, we limit our consideration to \mbox{$|n-m| \leqslant 1$}, selecting $\Delta t \approx \tau_{\text{IB}}$ so as to best satisfy both aforementioned conditions on $\Delta t$. The summation over $n$ and $m$ in \Eq{eq:Ksum} is therefore simplified to
\begin{equation}
\bar{K}(t) =  \delta_{i_N X}K_0 + \sum_{n=1}^{N-1} \delta_{i_n X}\left(K_0 + 2\delta_{i_{n+1} X}K_1\right)\label{eq:Kbar_NN}\,.
\end{equation}
Crucially, as shown in \App{sec:NN_example}, this reduction to a single summation allows us to re-express \Eq{eq:Pjk} as
\begin{equation}
P_{jk}(t) = e^{\delta_{j X} K_0} \sum_{i_{N-1}} \cdots \sum_{i_1} G_{i_N i_{N-1}} \cdots G_{i_2 i_1} M_{i_1 k}\,,\label{eq:Pjk2}
\end{equation}
where
\begin{equation}
G_{i_n i_{n-1}} = M_{i_n i_{n-1}}e^{\delta_{i_n X}(K_0 + 2\delta_{i_{n-1} X}K_1)}\,. \label{eq:Gpq}
\end{equation}
Equation~(\ref{eq:Pjk2}) can be compactly written in $2\!\times\!2$ matrix form in the $\ket{X}$, $\ket{C}$ basis:
\begin{equation}
\hat{P}(t) = \begin{pmatrix}
 P_{XX} & P_{XC}\\
 P_{CX} & P_{CC}
 \end{pmatrix} =
\begin{pmatrix}
 e^{K_0} & 0\\ 0 & 1
 \end{pmatrix}
 \hat{G}^{N-1}\hat{M}\label{eq:PNN}
\end{equation}
with $\hat{G}$ given by
\begin{equation}
\hat{G} = \begin{pmatrix}
 M_{XX}e^{K_0+2K_1} & M_{XC}\\ M_{CX}e^{K_0} & M_{CC} \label{eq:Gmatrix}
 \end{pmatrix}.
\end{equation}
It should be noted that our time step  $\Delta t \approx \tau_{\text{IB}}$ is too large to capture the initial rapid phonon-induced decay of the polarization associated with the phonon broadband~\cite{KrummheuerPRB02,MuljarovPRL04}. There is, however, a simple solution to this problem: for all $t<\tau_{\text{IB}}$, we replace our fixed $\Delta t$ with a variable $\Delta t'= t/2$. This ensures that $\bar{K}$ is calculated exactly for all $t<\tau_{\text{IB}}$. Further details on this modification are provided in \App{Sec:Cumulant}.

From the NN result \Eq{eq:PNN}, one can extract a simple analytic expression that describes the long-time behavior of the linear optical response. We use the asymptotic behavior of the standard IB model cumulant $K(t)$ in the long-time regime~\cite{KrummheuerPRB02,MuljarovPRL04},
\begin{equation}
K(t) \approx - i\Omega_p t - S\,, \label{eq:Kinf}
\end{equation}
where $\Omega_p$ is the polaron shift and $S$ is the Huang-Rhys factor (the explicit forms of which are provided in \App{Sec:IBLong}). This allows us to make the approximations $K_0 \approx - i\Omega_p \Delta t - S$ and $K_1 \approx S/2$. In the limit $\Delta t\approx  \tau_{\text{IB}} \ll \tau_{\text{JC}}$, this results in a fully analytic long-time dependence of the polarization (see \App{Sec:long} for further details):
\begin{equation}
\hat{P}(t) \approx e^{-\hat{S}/2}e^{-i\tilde{H}t}e^{-\hat{S}/2}\ \ \ \ (t> \tau_{\text{IB}}),
\label{eq:Panalyt}
\end{equation}
where
\begin{equation}
\tilde{H} = \begin{pmatrix}
\omega_X + \Omega_p & ge^{-S/2}\\
ge^{-S/2} & \omega_C
\end{pmatrix},
\ \ \ \
\hat{S} = \begin{pmatrix}
S & 0\\
0 & 0
\end{pmatrix}.
\label{eq:Htilde}
\end{equation}

Comparing the long-time analytics for $P_{jk}(t)$, given by Eqs.\,(\ref{eq:Panalyt}) and (\ref{eq:Htilde}), with the exact linear polarization in the JC model (no phonons),  $\bra{j}e^{-iH_{\text{JC}}t}\ket{k}$, we see that the effect of acoustic phonons in this limit ($\tau_{\text{IB}} \ll \tau_{\text{JC}}$) is a reduction of the exciton-cavity coupling strength $g$ by a factor of $e^{S/2}$ and the ZPL weight of the excitonic polarization by a factor of $e^{S}$. Additionally the bare exciton frequency is polaron-shifted: $\omega_X \to \omega_X + \Omega_p$. These facts are consistent with the analytic results of the IB model and are in agreement with previous experimental and theoretical works~\cite{Wilson-RaePRB02, WeiPRL14}. Furthermore, we note that the form of the modified Hamiltonian $\tilde{H}$ given by \Eq{eq:Htilde} is exactly the same as obtained after making the polaron transformation of the full Hamiltonian $H$. This work therefore provides a rigorous theoretical basis for taking this polaron transformed Hamiltonian as the unperturbed system in the widely used polaron master equation approaches~\cite{Wilson-RaePRB02,NazirJPCM16}.

We now address a general case in which the polaron and polariton time scales can be comparable, $\tau_{\text{IB}}\sim \tau_{\text{JC}}$, for example, in the case of a much larger exciton-cavity coupling $g$. This implies that we must find a way to reduce the time-step $\Delta t$ in the Trotter decomposition. We achieve this by going beyond the NN regime to the $L$-neighbor ($L$N) regime, where $L$ indicates the number of ``neighbors'' that we consider, corresponding to the condition \mbox{$|n-m| \leqslant L$} in \Eq{eq:Ksum}. The aforementioned condition $\Delta t \ll \tau_{\rm JC}$ applies equally to the $L$N regime, and therefore in this regime we are bound by the constraint $L \Delta t \gtrsim \tau_{\text{IB}}$. Importantly, this allows us to treat comparable polaron and polariton timescales provided that we choose $L$ such that the condition $\tau_{\rm IB}/L \ll \tau_{\rm JC}$ is satisfied.

In the $L$N approach we define a quantity $F_{i_L \cdots i_1}^{(n)}$ which is generated via a recursive relation
\begin{equation}
F_{i_L \cdots i_1}^{(n+1)} = \sum_{l=X,C} G_{i_L \cdots i_1 l} F_{i_{L-1} \cdots i_1 l}^{(n)}\,,
\label{Fs}
\end{equation}
using $F_{i_L \cdots i_1}^{(1)} = M_{i_1 k}$ as the initial value,
where $\hat{M}$ is defined as before by \Eq{eq:M}, while $G_{i_L \cdots i_1 l}$ is the $L$N  analog of \Eq{eq:Gpq}:
\begin{equation}
G_{i_L \cdots i_1 l} = M_{i_1 l}e^{\delta_{l X}(K_0 + 2\delta_{i_1 X}K_1 \cdots + 2\delta_{i_L X}K_L)}\,.
\label{Gtensor}
\end{equation}
The polarization is then given by
\begin{equation}
P_{jk}(t)  = e^{\delta_{j X} K_0}  F^{(N)}_{C\cdots C j}\,.
\label{eq:PLneighbours}
\end{equation}
Eqs. (\ref{Fs})\,--\,(\ref{eq:PLneighbours}) present an asymptotically exact solution for the linear polarization.  By extending the matrix size of the operators involved, it is straightforward to generalize this result to other correlators, such as the photon indistinguishability~\cite{GrangePRL15,HorneckerPRB17,Iles-SmithNatPhot17} or to other elements of the density matrix, such as the four-wave mixing polarization~\cite{KasprzakNatMat10,AlbertNatComm13}.

\begin{figure}[htp]
%\vskip-0.1cm
%\hskip-2.6cm
\centering
 \includegraphics[scale=0.42]{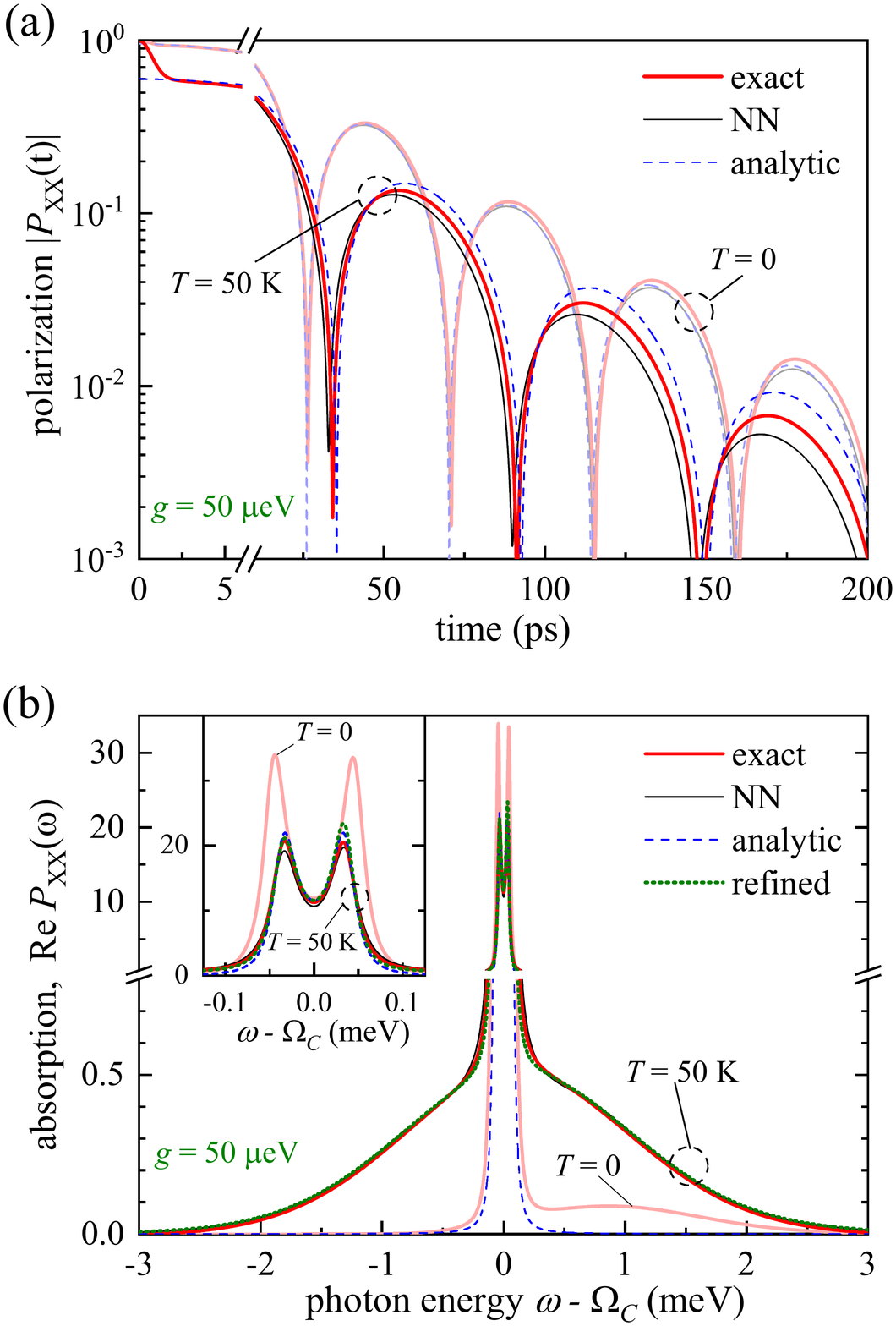}
%\vskip-2.4cm
%\hskip-1.8cm
%\includegraphics[scale=0.35]{Fig1b_new}
%\vskip-2.1cm
\caption{ (a) Excitonic linear polarization and (b) absorption for $T=0$ and 50\,K, calculated in the $L$N approach with $L=15$ (red thick solid lines), NN approach with $L=1$ (black thin solid lines), analytic approximation \Eq{eq:Panalyt} (blue dashed lines) and refined analytics (green dotted line). We use the realistic parameters of InGaAs QDs studied in~\cite{MuljarovPRL04,MuljarovPRL05} and
micropillars studied in~\cite{KasprzakNatMat10,AlbertNatComm13} (see also \App{Sec:Matrix_elements} for details) including $g=50\,\mu$eV, $\omega_X = \Omega_X -i\gamma_X$ with $\Omega_X=1329.6$\,meV and $\gamma_X = 2\,\mu$eV; $\omega_C = \Omega_C -i\gamma_C$ with $\Omega_C=\Omega_X + \Omega_p$, $\Omega_p = -49.8\,\mu$eV and $\gamma_C = 30\,\mu$eV.
Inset: linear plot of the absorption with limited frequency range.
}
\label{fig:P_small_g}
\end{figure}

To directly compare the various implementations of the Trotter decomposition method, we now apply the above-described formalisms to a system with realistic QD parameters~\cite{KasprzakNatMat10,AlbertNatComm13} in the regime of relatively small QD-cavity coupling ($g=50\,\mu$eV). Figure~\ref{fig:P_small_g}\,(a) shows the linear excitonic polarization $|P_{XX}(t)|$ calculated according to the analytic and NN techniques, Eqs.\,(\ref{eq:Panalyt}) and (\ref{eq:PNN}) respectively. Also shown is the ``exact'' polarization, calculated according the $L$-neighbor implementation, \Eq{eq:PLneighbours}, with $L=15$. In principle, one must take the limit $L \to \infty$ for a truly ``exact'' solution. For practical purposes, however, we select finite $L$ based on the desired accuracy; the 15-neighbor implementation provides a relative error in polarization of less than $0.1\%$ for the present set of parameters.

Figure~\ref{fig:P_small_g}\,(b) shows the excitonic absorption spectra for $g=50\,\mu$eV, calculated according to the above-described techniques. The absorption may be easily extracted from the linear polarization by taking the real part of the Fourier transform of ${P}_{XX}(t)$. The long-time behavior of the polarization is bi-exponential, as is clear from \Eq{eq:Panalyt}. The  absorption spectrum therefore consists of a well-resolved polariton doublet, described by the eigenvalues $\omega_j=\Omega_j - i\Gamma_j $ ($j=1,2$) of the effective Hamiltonian~\Eq{eq:Htilde}. Although not accounted for within the analytic model, there is a rapid initial decay in the polarization $|P_{XX}(t)|$; this short-time behavior correlates to the phonon broadband (BB) within the absorption spectrum. At lower temperatures, the BB is more asymmetric and the ZPL weight is increased, in agreement with the IB model. For the parameters selected and $T=50\,$K, $\tau_{\rm IB} \approx 3.2$\,ps and $\tau_{\rm JC}\approx \pi e^{S/2}/g\approx 57$\,ps  
(see Eqs. (\ref{eq:tauJC}) and (\ref{eq:tauIB}) alongside Appendices \ref{Sec:IBLong} and \ref{Sec:Matrix_elements}),
so that the NN approach presents a good approximation in this regime. As expected, the analytic result \Eq{eq:Panalyt} describes the long-time dynamics well but fails at short times, as it is clear from \Fig{fig:P_small_g}\,(a). This is manifested in the absorption spectrum in \Fig{fig:P_small_g}\,(b) as an absence of the BB. To improve on this shortcoming, we have additionally developed a {\it refined}, fully analytic solution (distinct from the above-described Trotter decomposition method) which captures the BB and reproduces the whole spectrum to very good accuracy in this regime, see the green dotted line in \Fig{fig:P_small_g}\,(b) and \App{Sec:refined} for details of the model.

In regimes of comparable polaron and polariton times $\tau_{\rm IB}\!\sim\! \tau_{\rm JC}$ (achieved by increasing the QD-cavity coupling constant to $g=0.6$\,meV while fixing all other parameters), the NN approach and the analytic approximations fail, leaving only the $L$N results. From the latter, we find that the long-time dynamics of the polarization matrix remain bi-exponential,
\begin{equation}
\hat{P}(t) \approx \sum_{j=1}^2 \hat{C}_j e^{-i\Omega_j t - \Gamma_j t}\ \ \ \ (t>\tau_{\rm IB})\,,
\label{eq:Pomega_pol}
\end{equation}
where $\Omega_j$ ($\Gamma_j$) are the polariton frequencies (linewidths) and $\hat{C}_j$ are the amplitude matrices.

\begin{figure}[htp]
%\vskip0.0cm
%\hskip-0.3cm
 \includegraphics[scale=0.42]{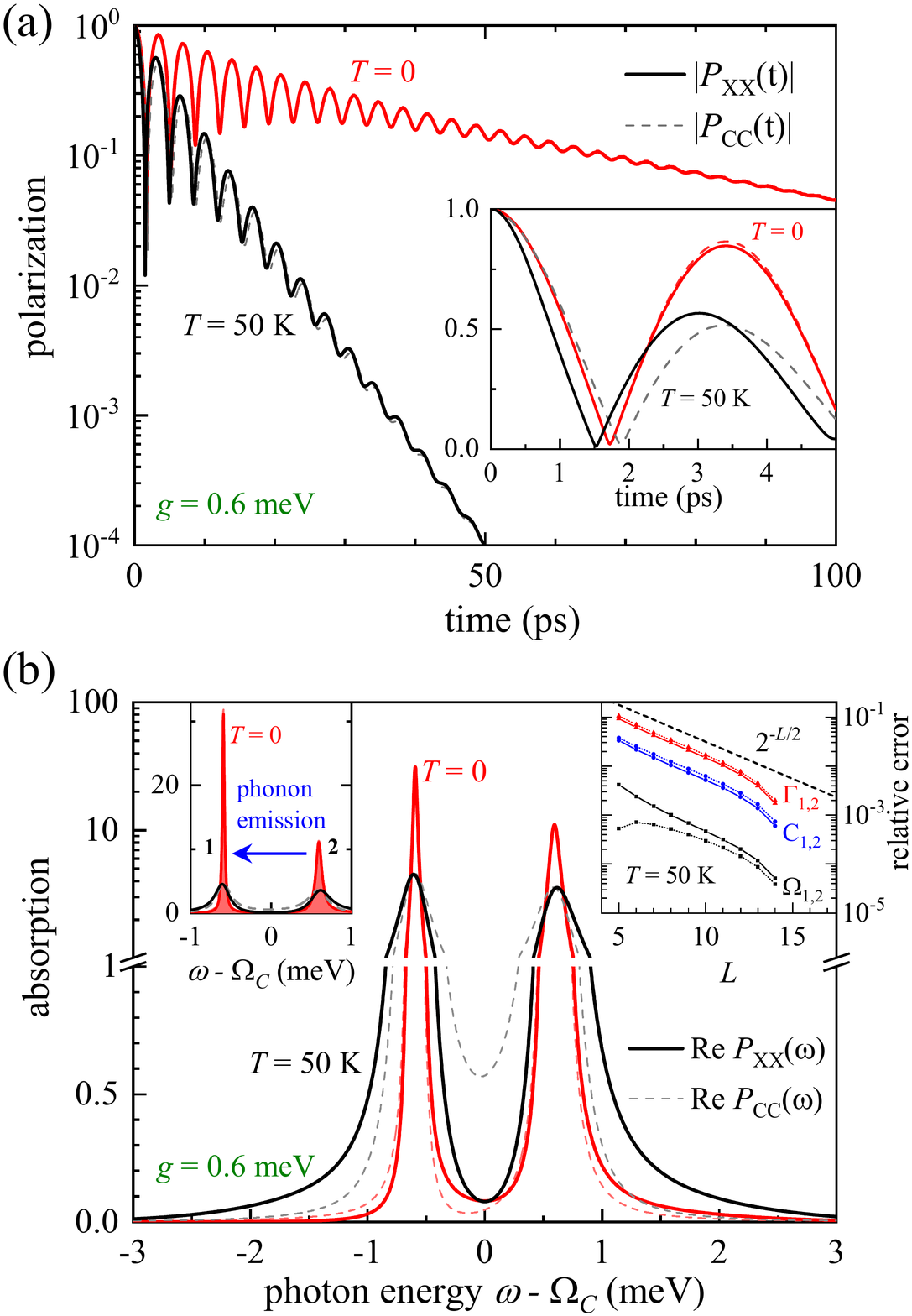}
%\vskip-2.4cm
%\hskip-2.2cm
%\includegraphics[scale=0.34]{Fig2b_new}
%\vskip-2.5cm
\caption{ As \Fig{fig:P_small_g} but for $g=0.6\,$meV and only $L$N result shown, for $T=0$ (red lines) and 50\,K (black lines). The photon polarization and absorption are also shown (dashed lines). Insets: (a) the initial polarization dynamics; (b, left) linear plot of the absorption illustrating the $2 \to 1$ polariton transition assisted by phonon emission; (b, right) the relative error for the parameters of the long-time bi-exponential dependence of $P_{XX}(t)$, \Eq{eq:Pomega_pol}, as a function of the number of neighbors $L$, taking $L=15$ as the exact solution.
\label{fig:P_large_g}
}
\end{figure}

The linear excitonic and cavity polarizations, $|P_{XX}(t)|$ and $|P_{CC}(t)|$, are shown in \Fig{fig:P_large_g}\,(a). There is a pronounced damping of the beating of the two exponentials, even for zero detuning (shown). This implies that the two peaks within the absorption spectra now have quite different linewidths, as is clear from \Fig{fig:P_large_g}\,(b).

\begin{figure}[htp]
\centering
%\vskip0.0cm
%\hskip-2.3cm
 \includegraphics[scale=0.34]{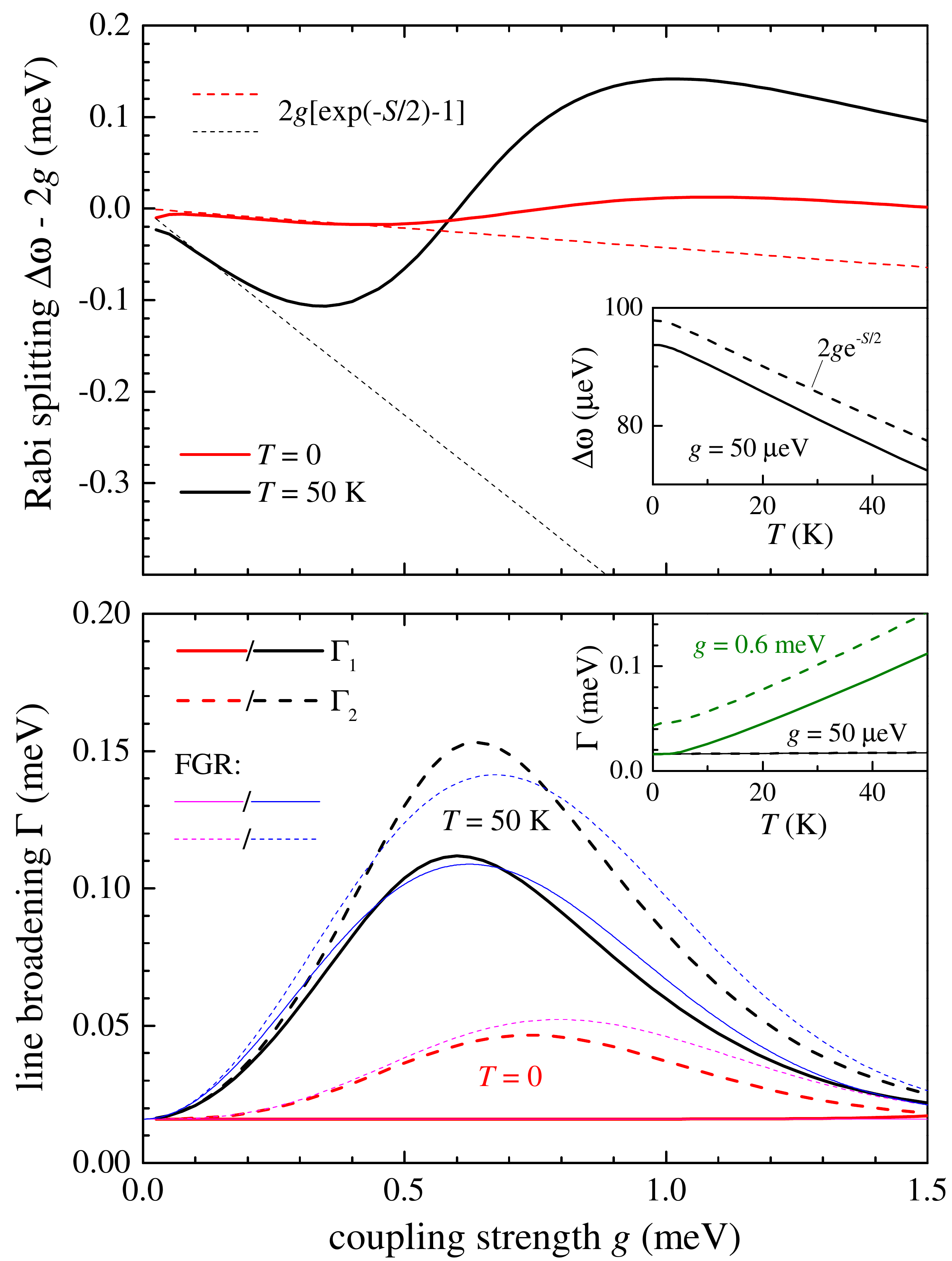}
%\vskip-2.1cm
%\vskip-0.4cm
% \includegraphics[scale=0.34]{Fig3}
%\vskip-0.6cm
\caption{Upper panel: Deviation of the polariton Rabi splitting $\Delta \omega = \Omega_2 - \Omega_1$, calculated via the $L$N model with $L=15$, from the nominal Rabi splitting $2g$ (solid lines), as a function of the exciton-cavity coupling strength $g$ for zero effective detuning, $\omega_C=\omega_X + \Omega_p$, and two different temperatures, $T=0$ and $T=50$\,K. The deviation of the phonon renormalized Rabi splitting from the nominal Rabi splitting $2g(e^{-S/2}-1)$ is shown by dashed lines. Inset in upper panel: the calculated full Rabi splitting $\Delta \omega$ (solid line) for $g=50\,\mu$eV as a function of the temperature $T$, in comparison with $2g e^{-S/2}$ (dashed lines). Lower panel: Linewidths $\Gamma_{1,2}$ of the lower (solid lines) and upper (dashed lines) polariton states in \Eq{eq:Pomega_pol} as functions of the coupling strength $g$, calculated in the $L$N approach with $L=15$ (thick black and red lines) and estimated according to Fermi's golden rule (thin blue and magenta lines). Inset in lower panel: temperature dependence of $\Gamma_{1,2}$ for $g=50\,\mu$eV (black) and 0.6\,meV (green).
\label{fig:Gamma_vs_g}
}
\end{figure}

The observed behavior can be understood in terms of real phonon assisted transitions between the states of the polariton doublet~\cite{MullerPRX15,Dory2016}. The variation in linewidths between $T=0$ K and $T = 50$ K shown in \Fig{fig:P_large_g}\,(b) is clear evidence of the phonon-induced broadening mechanism. At $T=0$, the high-energy polariton state (2) is significantly broader than the low-energy state (1) due to the allowed transition $2\to 1$, accompanied by emission of an acoustic phonon, as illustrated in the left inset of \Fig{fig:P_large_g}\,(b). At elevated temperatures both transitions $2\to 1$ and $1\to 2$,  with phonon emission and absorption respectively, are allowed, giving rise to more balanced linewidths. The line broadening as a function of temperature $T$ is shown in the inset of \Fig{fig:Gamma_vs_g}.

%The finite width of the low energy polariton state at $T=0$ is wholly due to the exciton dephasing $\gamma_X$ and radiative decay $\gamma_C$ inherit in the system.

Increasing the exciton-cavity coupling strength $g$ beyond $0.6$ meV (up to 1.5 meV), we find that the asymptotic behavior of the polarization retains the bi-exponential form of \Eq{eq:Pomega_pol}, thereby enabling direct comparison of polariton parameters at various coupling strengths $g$. The polariton line splitting $\Delta \omega = \Omega_2 - \Omega_1$ and linewidths $\Gamma_{1,2}$ are shown against $g$ in \Fig{fig:Gamma_vs_g}, whilst the behavior of the amplitude matrices $\hat{C}_{1,2}$ with $g$ is addressed in \App{sec:vary_g}.

The upper panel of \Fig{fig:Gamma_vs_g} shows the Rabi splitting $\Delta \omega$ of the polariton lines as a function of $g$, up to $g=1.5$ meV. In the regime of small $g$, the analytic calculation of Eqs.\,(\ref{eq:Panalyt}) and (\ref{eq:Htilde}) predict a phonon-renormalized Rabi splitting of $\Delta \omega  = 2g e^{-S/2}$ where $S$ is the Huang-Rhys factor defined in \App{Sec:IBLong}. This dependence is indeed observed in the 15-neighbor calculation for coupling strength $g$ below 0.2\,meV (0.5\,meV) for $T=50$\,K ($T=0$). A minor deviation from the analytic formula prediction of $\Delta \omega = 2g e^{-S/2}$ at small $g$ is due to finite exciton and cavity lifetimes used in the calculation: $\gamma_X = 2\,\mu$eV and $\gamma_C = 30\,\mu$eV. At larger $g$, the analytic prediction breaks down, and the Rabi splitting may even be enhanced by the presence of phonons.

The broadening $\Gamma_{1,2}$ of the polariton lines is strongly dependent on the exciton-cavity coupling strength $g$, as shown in the lower panel of \Fig{fig:Gamma_vs_g}. Maximal broadening occurs when the polariton splitting $\Delta \omega = \Omega_2 - \Omega_1$ corresponds to the typical energy of local acoustic phonons~\cite{MuljarovPRL05} (0.5\,--\,1\,meV for the QDs under consideration). To understand and quantify this behavior, we make a unitary transformation of the Hamiltonian $H = H_{\text{JC}} + H_{\text{IB}}$,
\begin{equation}
H \to H' = \hat{Y} H \hat{Y}^{-1}\,,
\end{equation}
where $\hat{Y}$ is the $2 \times 2$ matrix that diagonalizes the JC Hamiltonian $H_{\rm JC}$, comprising of diagonal elements $\alpha$ and off-diagonal elements $\pm \beta$ (see \App{Sec:long} for explicit forms of $\hat{Y}$, $\alpha$ and $\beta$). In making this transformation, we move from an exciton-cavity basis ($d^{\dagger}$, $a^{\dagger}$) to a polariton basis ($p_{1,2}^{\dagger}$). The transformed Hamiltonian $H'$ has the form,
\begin{equation}
H' = \begin{pmatrix} \omega_1 + \alpha^2 V & \alpha\beta V\\ \alpha\beta V & \omega_2 + \beta^2 V \end{pmatrix} + H_{\rm ph}\mathbb{1}\,,\label{eq:poltransmat}
\end{equation}
where $\omega_{1,2}$ are the eigenvalues of the JC Hamiltonian $H_{\rm JC}$ (see \App{Sec:long} for explicit forms), $V$ and $H_{\rm ph}$ are defined in \Eq{eq:HIB}, and $\mathbb{1}$ is a $2 \times 2$ identity matrix in the polariton basis. 

From \Eq{eq:poltransmat} it is clear that phonon assisted transitions between polariton states are permitted through the interaction term $\alpha\beta V(p_1^{\dagger}p_2 + p_2^{\dagger}p_1)$. Concentrating on this term, the contribution of real phonon-assisted transitions $\Gamma_{\rm ph}$ to the polariton broadening $\Gamma_{1,2}$ can be understood in terms of Fermi's golden rule (FGR)~\cite{MuljarovPRL05},
\be
\Gamma_{\rm ph} = \pi N_{\pm \Delta \omega/v_s} \sum_q |\alpha\beta\lambda_q|^2 \delta(\pm \Delta \omega - \omega_q)\,,\label{eq:Gammaif}
\ee
where $\lambda_q$ is the matrix element of the exciton-phonon coupling for the $q$-th phonon mode, $v_s$ is the speed of sound in the material, $\Delta \omega$ is the polariton Rabi splitting, and $N_{\pm \Delta \omega/v_s}$ is the Bose distribution function (\Eq{eq:BoseN}) evaluated at $q = \pm \Delta \omega/v_s$. We take the positive (negative) value of $\Delta \omega$ in \Eq{eq:Gammaif} for the $1 \rightarrow 2$ ($2 \rightarrow 1$) polariton transition.

Taking the average polariton Rabi splitting $\Delta \omega$ of $2g$ and approximating $\alpha$ and $\beta$ as $\alpha \approx \beta \approx \nicefrac{1}{\sqrt{2}}$ (valid in the case of zero detuning, or, more generally, in the regime $g \gg |\omega_X - \omega_C|$), we obtain the following expressions for the lower (1) and upper (2) polariton line broadenings,
\begin{align}
\Gamma_1 &= \Gamma_0 + N_{2g/v_s}\bar{\Gamma}_{\rm ph}\,,\label{eq:Gamma_FGR1}\\
\Gamma_2 &= \Gamma_0 + (N_{2g/v_s} + 1)\bar{\Gamma}_{\rm ph}\,,\label{eq:Gamma_FGR2}
\end{align}
where $\Gamma_0 = \nicefrac{1}{2}(\gamma_X + \gamma_C)$ is the intrinsic line broadening due to the long-time ZPL dephasing $\gamma_X$ and radiative decay $\gamma_C$, and, for a spherical Gaussian QD model (see \App{Sec:Matrix_elements}), $\bar{\Gamma}_{\rm ph}$ has the form
\begin{equation}
\bar{\Gamma}_{\rm ph} = \frac{g^3 (D_c - D_v)^2}{2\pi \rho_m v_s^5} \exp(-\frac{2 g^2 l^2}{v_s^2})\,.\label{eq:Gamma_FGRph}
\end{equation}

The linewidths $\Gamma_{1,2}$ calculated using Fermi's golden rule, Eqs.\,(\ref{eq:Gamma_FGR1}) and (\ref{eq:Gamma_FGR2}), are shown alongside the Trotter decomposition results in the lower panel of \Fig{fig:Gamma_vs_g}. There is, in general, remarkable agreement between Fermi's golden rule and the results obtained from the $L$N Trotter decomposition method; the small discrepancies may be attributed to multi-phonon transitions, which are not accounted for in FGR.

The inset in \Fig{fig:P_large_g}\,(b) demonstrates the quality of the present calculation at $g = 0.6$\,meV. For the values of $L$ shown, the error for the parameters of the long-time dependence \Eq{eq:Pomega_pol} decreases exponentially as $2^{-L/2}$. The computational time $t_{\rm c}$ is $\propto 2^L$, giving an error that scales as  $1/\sqrt{t_{\rm c}}$. Even for large $g$, the $L$N result quickly converges to the exact solution, with the relative error of the polariton linewidths $\Gamma_{1,2}$ saturating at a level below 1\%, as shown in \App{sec:vary_g}.
%$t_{\rm c}^{-\frac{1}{2}}$.

In conclusion, we have provided an asymptotically exact semi-analytic solution for the linear optical response of a QD-microcavity system coupled to an acoustic-phonon environment, valid for a wide range of system parameters.  Even for large cavity-QD coupling strength $g$, this solution reveals the dephasing mechanism in terms of real phonon-assisted transitions between polariton states of the Rabi doublet. For small $g$, our approach simplifies to an accurate analytic solution which provides an intuitive physical picture in terms of polaron-transformed polariton states superimposed with the phonon broadband, known from the independent boson model.

\begin{acknowledgments}
The authors acknowledge support by the EPSRC under the DTA
scheme and grant EP/M020479/1.
\end{acknowledgments}

\appendix
\section{Derivation of Eq.\,(\ref{eq:Pdef}) for the linear polarization\label{sec:compHamiltonian}}
We take as our starting point the standard definition of the optical polarization,
\begin{equation}
P = \text{Tr}\left\{\rho(t) c\right\}\,, \label{eq:genericP}
\end{equation}
where the annihilation operator $c$ stands either for the exciton operator $d$ or for the cavity operator $a$. Consequently, \Eq{eq:genericP} has the meaning of the full excitonic or photonic polarization, respectively.
Here $\rho(t)$ is the full density matrix of the system, including the exciton, cavity, and phonon degrees of freedom.

To obtain the {\em linear} polarization from \Eq{eq:genericP}, we first need to assume a pulsed excitation of the system at time $t=0$, which is described by the following evolution of the density matrix:
\begin{equation}
\rho(0_+) = e^{-i{\cal V}} \rho(-\infty) e^{i{\cal V}}, \label{eq:rho0+}
\end{equation}
where $\rho(-\infty)$ is the density matrix of a fully unexcited system, with its exciton-cavity part being in the absolute ground state $\ket{0}$ and phonons being in thermal equilibrium,
\begin{align}
\rho(-\infty) &= \ket{0} \bra{0} \rho_0\,,\\
\rho_0 &= e^{-\beta H_{\rm ph}} / \text{Tr}\left\{e^{-\beta H_{\rm ph}}\right\}_{\rm ph}\,.
\end{align}
Here, $\beta=(k_B T)^{-1}$, and the trace is taken over all possible phonon states.
The perturbation ${\cal V}$ due to the pulsed excitation has the form:
\begin{equation}
{\cal V} = \alpha(\tilde{c}^{\dagger} + \tilde{c}),
\end{equation}
where $\alpha$ is a constant, and again, $\tilde{c}$ is either $d$ or $a$, depending on the excitation (feeding) channel.

We assume that the evolution of the full density matrix of the exciton-cavity-phonon system after its optical pulsed excitation is given by the following standard Lindblad master equation
\begin{align}
i\dot{\rho}&=[{\cal H},\rho] + i\gamma_X\left(2d\rho d^{\dagger} - d^{\dagger}d\rho - \rho d^{\dagger} d\right)\nonumber\\
&+ i\gamma_C\left(2a\rho a^{\dagger} - a^{\dagger}a\rho - \rho a^{\dagger} a\right),
\end{align}
in which the Hamiltonian ${\cal H}={\cal H}_{\rm JC}+H_{IB}$ is {\it Hermitian}. Here, ${\cal H}_{\rm JC}$ is the JC Hamiltonian $H_{\rm JC}$ defined by Eq.\,(1) in which the complex frequencies
\begin{equation}
\omega_{X,C} =\Omega_{X,C} - i\gamma_{X,C}\,, \hspace{1cm} \Omega_{X,C}, \gamma_{X,C} \in \mathbb{R}\,,
\end{equation}
are replaced by real ones by removing the imaginary parts: $\omega_{X,C}  \rightarrow \Omega_{X,C}$. Noting that
$$%\be
[{\cal H},\rho] = H\rho - \rho H^\ast + i\gamma_X(d^{\dagger}d\rho + \rho d^{\dagger} d) + i\gamma_C(a^{\dagger}a\rho + \rho a^{\dagger} a)\,,
$$%\ee
where $H$ is the full {\it non-Hermitian} Hamiltonian defined on the first page of the main text and $H^\ast$ is its complex conjugate, we may re-express the Lindblad master equation as
\be
i\dot{\rho}=H\rho-\rho H^\ast +2i\gamma_Xd\rho d^\dagger+2i\gamma_C a\rho a^\dagger\,.
\label{Lindblad}
\ee

In the linear polarization, we keep in the full polarization only the terms which are linear in $\alpha$. Looking closer, this implies keeping only $\ket{X} \bra{0}$ and $\ket{C} \bra{0}$ elements of the density matrix. When the density matrix is reduced to only $\ket{X} \bra{0}$ and $\ket{C} \bra{0}$ elements, the last two terms in \Eq{Lindblad} vanish, which yields an explicit solution:
\be
\rho(t)=e^{-iHt}\rho(0_+)e^{iH^\ast t}\,,
\ee
in which $H^\ast$ can actually be replaced by $H_{\rm ph}$. The linear polarization then takes the form
\be
P_L(t)=-i\alpha \text{Tr}\left\{ e^{-iHt} \tilde{c}^{\dagger}  \ket{0} \bra{0} \rho_0 e^{iH_{\rm ph} t} c \right\}
\ee
Now, dropping the unimportant constant factor $-i\alpha$ and introducing indices $j,k=X,C$ to replace the operators $\tilde{c}^{\dagger}$ and $c$, we arrive at Eq.\,(\ref{eq:Pdef}) of the
main text.

\section{Trotter decomposition of the evolution operator\label{sec:trotterU}}

Using the Trotter decomposition, the evolution operator is presented in \Eq{eq:Trotter} as $\hat{U}(t)=\lim_{N\to\infty} \hat{U}_N(t)$, where
\begin{align}
\hat{U}_N(t) &=  e^{iH_{\rm ph}t} e^{-i H_{\rm IB} (t-t_{N-1})} e^{-i H_{\rm JC} (t-t_{N-1})}
\cdots \nonumber\\
&\hspace{-6mm}\times e^{-i H_{\rm IB} (t_n-t_{n-1})} e^{-i H_{\rm JC} (t_n-t_{n-1})}
\cdots \nonumber\\
&\hspace{-6mm}\times e^{-i H_{\rm IB} t_1} e^{-i H_{\rm JC} t_1} \nonumber
\\
&\hspace{-9mm}= e^{iH_{\rm ph}t} e^{-i H_{\rm IB} (t-t_{N-1})} e^{-iH_{\rm ph}t_{N-1}} e^{-i H_{\rm JC} (t-t_{N-1})}
\cdots
\nonumber
\\
&\hspace{-6mm}\times e^{iH_{\rm ph}t_n} e^{-i H_{\rm IB} (t_n-t_{n-1})} e^{-iH_{\rm ph}t_{n-1}} e^{-i H_{\rm JC} (t_n-t_{n-1})}
\cdots \nonumber\\
&\hspace{-6mm}\times e^{iH_{\rm ph}t_1} e^{-i H_{\rm IB} t_1} e^{-i H_{\rm JC} t_1}
\nonumber\\
&\hspace{-9mm}= \hat{W}(t,t_{N-1})\hat{M}(t- t_{N-1}) \cdots\nonumber\\
&\hspace{-6mm}\times \hat{W}(t_n,t_{n-1})\hat{M}(t_n - t_{n-1}) \cdots \hat{W}(t_1,0)\hat{M}(t_1)
\,,
\label{eq:Trotter2}
\end{align}
where we have used the fact that the operators $H_{\rm ph}$ and $H_{\rm JC}$ commute. From the definition of $H_{\rm IB}$ we note that
\be
\hat{W}(t_n,t_{n-1}) =e^{iH_{\rm ph}t_n} e^{-i H_{\rm IB} (t_n-t_{n-1})} e^{-iH_{\rm ph}t_{n-1}}
\ee
is a diagonal operator in the $2$-basis state matrix representation in terms of $\ket{X}$  and $\ket{C}$:
\be
\hat{W}(t_n,t_{n-1})=\begin{pmatrix} W_X(t_n,t_{n-1}) &0\\0 & W_C(t_n,t_{n-1})\end{pmatrix}
\ee
with
\begin{align*}
W_X(t_n,t_{n-1}) &= e^{iH_{\rm ph}t_n} e^{-i (H_{\rm ph}+V) (t_n-t_{n-1})} e^{-iH_{\rm ph}t_{n-1}},\\
W_C(t_n,t_{n-1}) &= 1.
\end{align*}
Using the time ordering operator $ \mathcal{T}$, $\hat{W}$-matrix element $W_X$ can be written as
\be
W_X(t_n,t_{n-1}) =\mathcal{T} \exp{-i\int_{t_{n-1}}^{t_n} V(\tau) d\tau},
\ee
where $V(\tau) = e^{iH_{\rm ph}\tau}Ve^{-iH_{\rm ph}\tau}$ is the interaction representation of the exciton-phonon coupling $V$, which is given by \Eq{eq:HIB} of the main text.

Substituting the evolution operator \Eq{eq:Trotter2} into \Eq{eq:Pdef} for the polarization $P_{jk}(t) $ and explicitly expressing the matrix products gives
\begin{align}
P_{jk}(t) &=  \sum_{i_{N-1} = X,C} \cdots \sum_{i_1=X,C} \langle W_{i_N}M_{i_N i_{N-1}}\nonumber\\
&\hspace{5mm}\times W_{i_{N-1}}M_{i_{N-1}i_{N-2}} \cdots M_{i_{n+1}i_n}W_{i_n}M_{i_n i_{n-1}} \cdots\nonumber\\
&\hspace{5mm}\times W_{i_1}M_{i_1 i_0} \rangle_{\rm ph}
\label{eq:Pjk_timeordered}
\end{align}
with $i_N=j$ and $i_0=k$. From here, we note that only $W$ elements contain the phonon interaction and through a simple rearrangement of \Eq{eq:Pjk_timeordered} we arrive at \Eq{eq:Pjk} of the main text.

\section{Linear polarization in the NN approximation, including an example realization\label{sec:NN_example}}
The single summation in the cumulant Eq. (\ref{eq:Kbar_NN}) allows us to express, for each realization, the expectation value in Eq.\,(\ref{eq:Pjk}) as a product
\begin{align}
&\langle W_{i_N}(t,t_{N-1})\cdots W_{i_n}(t_n,t_{n-1}) \cdots W_{i_2}(t_2,t_1) W_{i_1}(t_1,0)\rangle_{\rm ph} \nonumber\\
&\hspace{5mm}=e^{\delta_{i_N X}K_0} e^{\delta_{i_{N-1} X}\left(K_0 + 2\delta_{i_N X}K_1\right)} \cdots\nonumber\\
&\hspace{10mm}\times e^{\delta_{i_{n-1} X}\left(K_0 + 2\delta_{i_n X}K_1\right)} \cdots e^{\delta_{i_{1} X}\left(K_0 + 2\delta_{i_2 X}K_1\right)}\label{eq:exp_product}.
\end{align}
It is convenient to introduce
\begin{equation}
R_{i_n i_{n-1}} = e^{\delta_{i_{n-1} X}\left(K_0 + 2\delta_{i_n X}K_1\right)},
\end{equation}
enabling us to express the expectation values of the product of W-operators for a given realization \Eq{eq:exp_product} as $e^{\delta_{i_N X}K_0}R_{i_N i_{N-1}} \cdots  R_{i_2 i_1}$. Inserting this expression into \Eq{eq:Pjk}, we  find
\begin{align}
P_{jk}(t) &= e^{\delta_{i_N X}K_0} \sum_{i_{N-1} = X,C} \cdots \sum_{i_1=X,C}\hspace{40mm} \nonumber\\
&\hspace{-5mm}\left(M_{i_N i_{N-1}} \cdots M_{i_2 i_1} M_{i_1 i_0}\right) \left(R_{i_N i_{N-1}} \cdots R_{i_2 i_1}\right).
\label{Pjk}
\end{align}
We then join together corresponding $M_{i_n i_{n-1}}$ and $R_{i_n i_{n-1}}$ elements through the definition of a matrix
\begin{equation}
G_{i_n i_{n-1}} =  M_{i_n i_{n-1}} R_{i_n i_{n-1}}\,,
\end{equation}
%$G_{i_n i_{n-1}} =  M_{i_n i_{n-1}} R_{i_n i_{n-1}}$
which transforms \Eq{Pjk} to
\begin{align}
P_{jk}(t) &= e^{\delta_{i_N X}K_0} \sum_{i_{N-1} = X,C} \cdots \sum_{i_{n-1} = X,C} \cdots \sum_{i_1=X,C}\nonumber\\
&G_{i_N i_{N-1}} \cdots G_{i_n i_{n-1}} \cdots G_{i_2 i_1} M_{i_1 i_0}\,.
\label{Pjkt}
\end{align}
Using the fact that $i_N=j$ and $i_0=k$, we arrive at \Eq{eq:Pjk2} which is compactly represented in \Eq{eq:PNN} as a product of matrices.

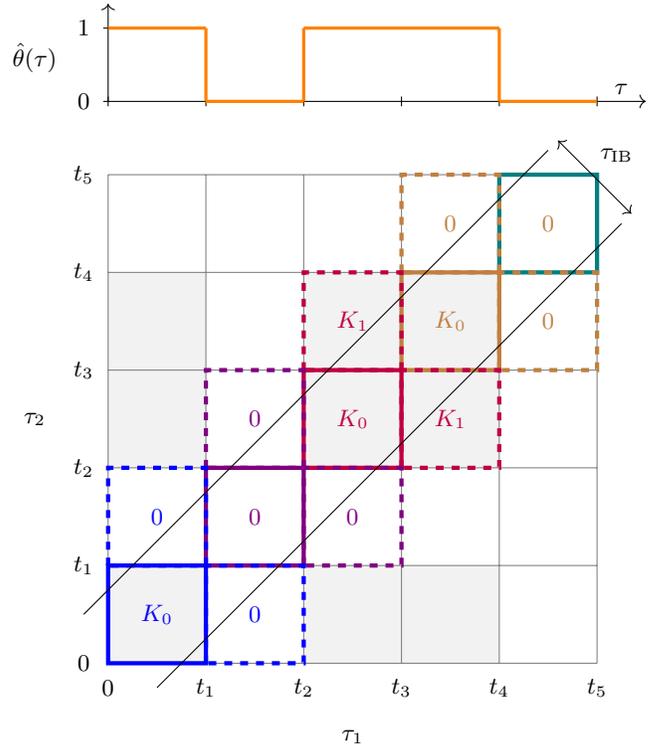
\begin{figure}[htp]
\centering
\begin{tikzpicture}[scale=0.65]
    \fill[gray!10](0,0) rectangle (2,2);
    \fill[gray!10](0,4) rectangle (2,8);
    \fill[gray!10](4,0) rectangle (8,2);
    \fill[gray!10](4,4) rectangle (8,8);
    \draw[step=2cm,color=gray] (0, 0) grid (10, 10);
    %\draw[very thick,step=2cm] (0, 0) grid (2, 2);
    %\node at (1,1) {$K_0 (\Delta t)$};

    \draw[ultra thick,color=teal](8,8) rectangle (10,10);
    \node at (9,9) {\color{brown} 0};

    \draw[ultra thick,color=brown](6,6) rectangle (8,8);
    \node at (7,7) {\color{brown} $K_0$};
    \draw[ultra thick,color=brown,dashed](6,8) rectangle (8,10);
    \draw[ultra thick,color=brown,dashed](8,6) rectangle (10,8);
    \node at (7,9) {\color{brown} 0};
    \node at (9,7) {\color{brown} 0};

    \draw[ultra thick,color=purple](4,4) rectangle (6,6);
    \node at (5,5) {\color{purple} $K_0$};
    \draw[ultra thick,color=purple,dashed](4,6) rectangle (6,8);
    \draw[ultra thick,color=purple,dashed](6,4) rectangle (8,6);
    \node at (5,7) {\color{purple} $K_1$};
    \node at (7,5) {\color{purple} $K_1$};

    \draw[ultra thick,color=violet](2,2) rectangle (4,4);
    \node at (3,3) {\color{violet} 0};
    \draw[ultra thick,color=violet,dashed](4,2) rectangle (6,4);
    \draw[ultra thick,color=violet,dashed](2,4) rectangle (4,6);
    \node at (3,5) {\color{violet} 0};
    \node at (5,3) {\color{violet} 0};

    \draw[ultra thick,color=blue](0,0) rectangle (2,2);
    \node at (1,1) {\color{blue}$K_0$};
    \draw[ultra thick,color=blue,dashed](2,0) rectangle (4,2);
    \draw[ultra thick,color=blue,dashed](0,2) rectangle (2,4);
    \node at (1,3) {\color{blue} 0};
    \node at (3,1) {\color{blue} 0};
    \node at (0,-0.5) {0};
    \node at (2,-0.5) {$t_1$};
    \node at (4,-0.5) {$t_2$};
    \node at (6,-0.5) {$t_3$};
    \node at (8,-0.5) {$t_4$};
    \node at (10,-0.5) {$t_5$};
    \node at (-0.5,0) {0};
    \node at (-0.5,2) {$t_1$};
    \node at (-0.5,4) {$t_2$};
    \node at (-0.5,6) {$t_3$};
    \node at (-0.5,8) {$t_4$};
    \node at (-0.5,10) {$t_5$};
    \node at (5,-1.5) {$\tau_1$};
    \node at (-1.5,5) {$\tau_2$};

    \draw (-0.5,1) -- (9,10.5);
    \draw (1,-0.5) -- (10.5,9);
    \draw [<->] (9.2,10.7) -- (10.7,9.2);
    \node at (10.4,10.4) {$\tau_{\rm IB}$};

    \draw [->] (-0.1,11.5) -- (11,11.5);
    \draw [->] (0,11.5) -- (0,13.5);
    \draw (-0.1,13) -- (0.1,13);
    \node at (10.5,11.75) {$\tau$};
    \node at (-1.5,12.4) {$\hat{\theta}(\tau)$};
    \node at (-0.5,11.5) {0};
    \node at (-0.5,13) {1};
    \draw (0,11.4) -- (0,11.6);
    \draw (2,11.4) -- (2,11.6);
    \draw (4,11.4) -- (4,11.6);
    \draw (6,11.4) -- (6,11.6);
    \draw (8,11.4) -- (8,11.6);
    \draw (10,11.4) -- (10,11.6);

    \draw [very thick,color=orange] (0,13) -- (2,13);
    \draw [very thick,color=orange] (2,13) -- (2,11.5);
    \draw [very thick,color=orange] (2,11.5) -- (4,11.5);
    \draw [very thick,color=orange] (4,11.5) -- (4,13);
    \draw [very thick,color=orange] (4,13) -- (8,13);
    \draw [very thick,color=orange] (8,13) -- (8,11.5);
    \draw [very thick,color=orange] (8,11.5) -- (10,11.5);

\end{tikzpicture}
\caption{ Example realization for the NN implementation with $N=5$. In this realization, $i_1 = X$, $i_2 = C$, $i_3 = X$, $i_4 = X$, $i_5 = C$, as is clear from the step function $\hat{\theta}(t)$ associated with the given realization, shown on the top. \label{fig:grid}}
\end{figure}

To illustrate this idea by way of an example, we take a particular realization for $N=5$, provided for illustration in \Fig{fig:grid}. In this realization, $i_1 = X$, $i_2 = C$, $i_3 = X$, $i_4 = X$, and $i_5 = C$.
Each exponential $e^{\delta_{i_{n-1} X}\left(K_0 + 2\delta_{i_n X}K_1\right)}$ in \Eq{eq:exp_product} can be visualized as an L-shaped portion of the time grid (color coded in the figure). In the illustrated realization we have,
\begin{align*}
R_{i_2 i_1} = e^{\delta_{i_{1} X}\left(K_0 + 2\delta_{i_2 X}K_1\right)} &= e^{K_0}\,,\\
R_{i_3 i_2} =e^{\delta_{i_{2} X}\left(K_0 + 2\delta_{i_3 X}K_1\right)} &= e^0 = 1\,,\\
R_ {i_4 i_3}=e^{\delta_{i_{3} X}\left(K_0 + 2\delta_{i_4 X}K_1\right)} &= e^{K_0 + 2K_1}\,,\\
R_ {i_5 i_4}=e^{\delta_{i_{4} X}\left(K_0 + 2\delta_{i_5 X}K_1\right)} &= e^{K_0}\,,\\
e^{\delta_{i_5 X} K_0} &= e^0 = 1\,.
\end{align*}
We then find
\begin{align*}
G_{i_2 i_1} &= G_{CX} = M_{CX}e^{K_0}\,,\\
G_{i_3 i_2} &= G_{XC} = M_{XC}\,,\\
G_{i_4 i_3} &= G_{XX} = M_{XX}e^{K_0 + 2K_1}\,,\\
G_{i_5 i_4} &= G_{CX} = M_{CX}e^{K_0}\,,
\end{align*}
which contributes to the total polarization \Eq{Pjkt}.

Note that the condition for the NN approximation to be valid is also illustrated in \Fig{fig:grid}: All the time moments of integration for which $|\tau_2-\tau_1|<\tau_{\rm IB}$ should be located within the colored squares, which are taken into account in the NN calculation of the cumulant.

\section{Calculation of $K_{|n-m|}$ from the IB model cumulant}
\label{Sec:Cumulant}

\begin{figure}[htp]
%\begin{subfigure}[b]{0.4\textwidth}
\begin{tikzpicture}[scale=0.9]
    \fill[gray!20](0,0) rectangle (2,2);
    \draw[step=2cm,color=gray] (0, 0) grid (6, 6);
    \draw[very thick,step=2cm] (0, 0) grid (2, 2);
    \node at (-1.5,6) {(a)};
    \node at (1,1) {$K_0 (\Delta t)$};
    \node at (0,-0.5) {0};
    \node at (2,-0.5) {$t_1$};
    \node at (4,-0.5) {$t_2$};
    \node at (6,-0.5) {$t_3$};
    \node at (-0.5,0) {0};
    \node at (-0.5,2) {$t_1$};
    \node at (-0.5,4) {$t_2$};
    \node at (-0.5,6) {$t_3$};
    \node at (3,-1) {$\tau_1$};
    \node at (-1,3) {$\tau_2$};
\end{tikzpicture}

\vspace{0.6cm}
%\caption{Graphical representation of the use of the IB model cumulant $K(t)$ for finding $K_0$. \label{fig:K0}}
%\end{subfigure}
%
%
%\begin{subfigure}[b]{0.4\textwidth}
\begin{tikzpicture}[scale=0.9]
    \fill[gray!20](0,0) rectangle (2,2);
    \fill[gray!20](2,2) rectangle (4,4);
    \fill[lightgray!20](0,2) rectangle (2,4);
    \fill[lightgray!20](2,0) rectangle (4,2);
    \draw[step=2cm,color=gray] (0, 0) grid (6, 6);
    \draw[very thick,step=4cm] (0, 0) grid (4, 4);
    \node at (-1.5,6) {(b)};
    \node at (1,1) {$K_0 (\Delta t)$};
    \node at (1,3) {$K_1 (\Delta t)$};
    \node at (3,1) {$K_1 (\Delta t)$};
    \node at (3,3) {$K_0 (\Delta t)$};
    \node at (0,-0.5) {0};
    \node at (2,-0.5) {$t_1$};
    \node at (4,-0.5) {$t_2$};
    \node at (6,-0.5) {$t_3$};
    \node at (-0.5,0) {0};
    \node at (-0.5,2) {$t_1$};
    \node at (-0.5,4) {$t_2$};
    \node at (-0.5,6) {$t_3$};
    \node at (3,-1) {$\tau_1$};
    \node at (-1,3) {$\tau_2$};
\end{tikzpicture}
%\caption{As \Fig{fig:Knm}(a) but for $K_1$. \label{fig:K1} \vspace{0.36cm}}
%\end{subfigure}
\caption{Graphical representations of the use of the IB model cumulant $K(t)$ for finding (a) $K_{0}(\Delta t)$ and (b) $K_{1}(\Delta t)$.}
\label{fig:Knm}
\end{figure}
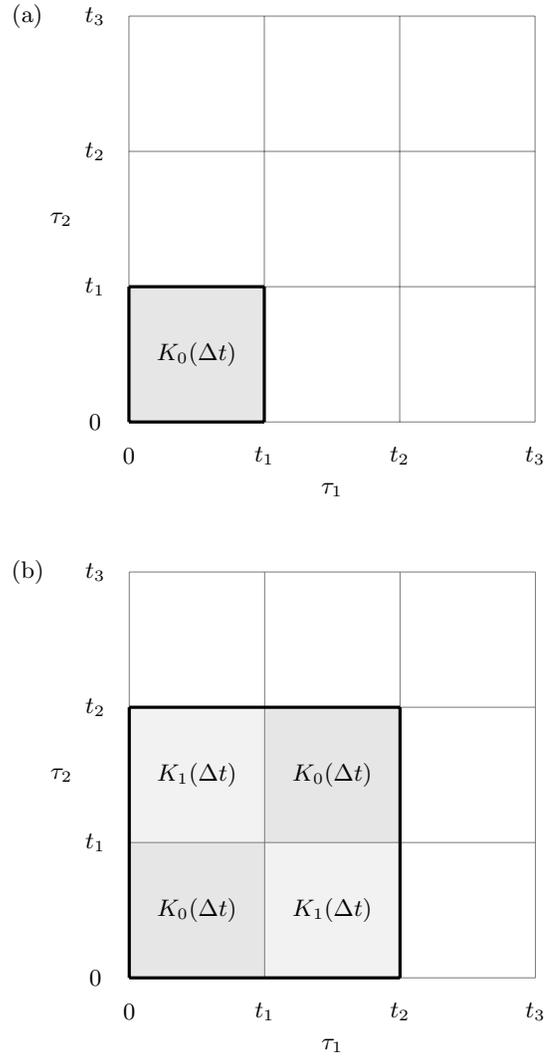

As is clear from the definition given in \Eq{eq:Knm} of the main text, the integral $K_{|n-m|}$ depends only on the difference $|n-m|$; it is depicted graphically in \Fig{fig:Knm} (a). To find $K_0$, we set $m=n=1$,
\be
K_{0} = - \frac{1}{2} \int_{0}^{t_1} d\tau_1 \int_{0}^{t_1} d\tau_2 \langle \mathcal{T} V(\tau_1)V(\tau_2)\rangle= K(\Delta t)\,,
\ee
where $K(t)$ is the IB cumulant, which is calculated explicitly in \App{Sec:IBLong} below, see Eq. (\ref{eq:fullK}). \\
\\
Analogously, to find $K_1$ we may set $m=1$ and $n=2$ which gives
\begin{equation}
K_{1} = - \frac{1}{2} \int_{t_1}^{t_2} d\tau_1 \int_{0}^{t_1} d\tau_2 \langle \mathcal{T} V(\tau_1)V(\tau_2)\rangle\label{eq:K1a} \,,
\end{equation}
or, by setting $m=2$ and $n=1$ instead, we obtain the same result:
\begin{equation}
K_{1} = - \frac{1}{2} \int_{0}^{t_1} d\tau_1 \int_{t_1}^{t_2} d\tau_2 \langle \mathcal{T} V(\tau_1)V(\tau_2)\rangle\label{eq:K1b} \,.
\end{equation}
Eqs. (\ref{eq:K1a}) and (\ref{eq:K1b}) correspond to the squares labeled as $K_1$ in \Fig{fig:Knm} (b). In order to calculate $K_1$ from the IB cumulant, we note that
\begin{equation}
K(2\Delta t) = 2K_0 + 2K_1.
\end{equation}
Therefore,
\begin{equation}
K_1 = \frac{1}{2}\left[K(2\Delta t) - 2K_0\right].
\end{equation}
In general, all the integrals $K_p$ can be found recursively:
\begin{align}
K_{p>0} &= \frac{1}{2}\Bigg[K((p+1)\Delta t) - (p+1)K_0\nonumber\\
&\hspace{5mm} - \sum_{q=1}^{p-1} 2(p+1-q)K_q \Bigg].
\end{align}

%\vspace{1cm}
%\hspace{-1.5cm}
\begin{figure}[htp]
\begin{tikzpicture}[scale=0.675]
    \fill[teal!10](0,0) rectangle (3,3);
    \fill[teal!10](3,3) rectangle (6,6);
    \fill[teal!5](0,3) rectangle (3,6);
    \fill[teal!5](3,0) rectangle (6,3);
    %\fill[gray!10](0,0) rectangle (6,6);
    \draw[step=4cm,color=gray] (0, 0) grid (8, 8);
    \draw[very thick,color=teal,step=3cm] (0, 0) grid (6, 6);
    %\node at (-1.5,8) {{\large (c)}};
    \node at (0,-0.5) {\color{teal}0};
    \node at (3,-0.5) {\color{teal} $\Delta t'$};
    \node at (6,-0.5) {\color{teal} $2\Delta t'$};
    \node at (-0.5,0) {\color{teal}0};
    \node at (-0.5,3) {\color{teal} $\Delta t'$};
    \node at (-0.5,6) {\color{teal} $2\Delta t'$};
    \node at (1.5,1.5) {\color{teal} $K_0(\Delta t')$};
    \node at (1.5,4.5) {\color{teal} $K_1(\Delta t')$};
    \node at (4.5,1.5) {\color{teal} $K_1(\Delta t')$};
    \node at (4.5,4.5) {\color{teal} $K_0(\Delta t')$};
\end{tikzpicture}
\caption{Adaptation of the grid of \Fig{fig:Knm} for small time: $t < \tau_{\rm IB}$. The grey grid illustrates the $\Delta t$ discretization used for $t > \tau_{\rm IB}$ (as shown in \Fig{fig:Knm}), whilst the green grid illustrates the adapted discretization for $t < \tau_{\rm IB}$. In this small time regime and the $L=1$ implementation, a $2 \times 2$ grid is always used, giving $\Delta t' = t/2$. More generally, the $L$N implementation requires a grid of size $(L+1)\times(L+1)$ for $t < \tau_{\rm IB}$. \label{fig:Kgrid_smallt}}
\end{figure}
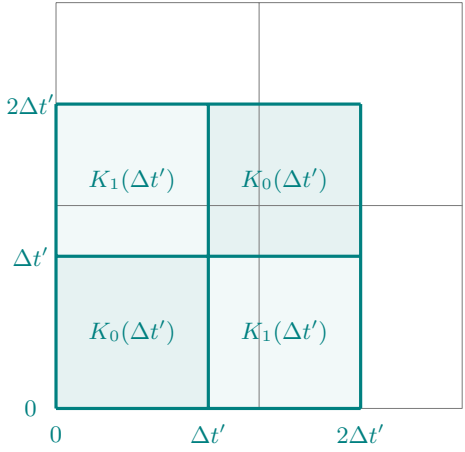

For all $t < \tau_{\rm IB}$, we modify our approach by replacing our fixed $\Delta t$ with variable $\Delta t' =  t/(L+1)$, where $L$ is the chosen number of neighbors. Accordingly, in this regime time is discretized into $L+1$ tranches. For example, the NN ($L=1$) approach uses a $2 \times 2$ grid, as shown in \Fig{fig:Kgrid_smallt}. Crucially, this ensures that no portions of the $K(t)$ grid are neglected. We therefore may allow $\Delta t'$ to become arbitrarily small whilst always exactly calculating $K(t)$.  Note that this is only valid for $t < \tau_{\rm IB}$: If we were to extend this approach to $t > \tau_{\rm IB}$ then for some values of $t$ our time interval $\Delta t'$ would become too large, and the accuracy of the calculation would be degraded.

%\begin{figure}[h]
%\centering
%\begin{tikzpicture}[scale=0.7]
%    \fill[teal!10](0,0) rectangle (3,3);
%    \fill[teal!10](3,3) rectangle (6,6);
%    \fill[teal!5](0,3) rectangle (3,6);
%    \fill[teal!5](3,0) rectangle (6,3);
%    %\fill[gray!10](0,0) rectangle (6,6);
%    \draw[step=4cm,color=gray] (0, 0) grid (8, 8);
%    \draw[very thick,color=teal,step=3cm] (0, 0) grid (6, 6);
%    \node at (0,-0.5) {\color{teal}0};
%    \node at (3,-0.5) {\color{teal} $\Delta t'$};
%    \node at (6,-0.5) {\color{teal} $2\Delta t'$};
%    \node at (-0.5,0) {\color{teal}0};
%    \node at (-0.5,3) {\color{teal} $\Delta t'$};
%    \node at (-0.5,6) {\color{teal} $2\Delta t'$};
%    \node at (1.5,1.5) {\color{teal} $K_0(\Delta t')$};
%    \node at (1.5,4.5) {\color{teal} $K_1(\Delta t')$};
%    \node at (4.5,1.5) {\color{teal} $K_1(\Delta t')$};
%    \node at (4.5,4.5) {\color{teal} $K_0(\Delta t')$};
%\end{tikzpicture}
%\caption{The adaptation of the grid of \Fig{fig:Knm} for small time: $t < \tau_{\rm IB}$. The grey grid illustrates the $\Delta t$ discretisation used for $t > \tau_{\rm IB}$ (as for \Fig{fig:Knm}), whilst the green grid illustrates the adapted discretisation for $t < \tau_{\rm IB}$. In this small time regime, the time $t$ is always split into a $2 \times 2$ grid, giving $\Delta t' = t/2$.\label{fig:Kgrid_smallt}}
%\end{figure}

\section{The IB model cumulant and its long-time behavior, Eq.\,(\ref{eq:Kinf})}
\label{Sec:IBLong}

The IB model cumulant $K(t) $ can be conveniently written in terms of the standard phonon propagator $D_q$~\cite{MuljarovPRL04},
\be
K(t)= - \frac{i}{2} \int_0^t d\tau_1 \int_0^t d\tau_2 \sum_q |\lambda_q|^2 D_q(\tau_1 - \tau_2)\,,
\label{Cum}
\ee
where
\begin{align}
i D_q(t) &= \langle \mathcal{T} [b_q(t)+b^\dagger_{-q}(t)]^\dagger [b_q(0)+b^\dagger_{-q}(0)]\rangle\nonumber\\
&=N_q e^{i\omega_q |t|} + (N_q + 1) e^{-i\omega_q |t|}
\label{Ddef}
\end{align}
and $N_q$ is the Bose distribution function,
\begin{equation}
N_q = \frac{1}{e^{\beta \omega_q} - 1}.\label{eq:BoseN}
\end{equation}
Performing the integration in \Eq{Cum}, we obtain
\begin{align}
K(t) &= \sum_q |\lambda_q|^2 \left(\frac{N_q}{\omega_q^2}\left[e^{i\omega_q t} - 1\right]\right. \nonumber\\
&\hspace{20mm}\left. + \frac{N_q + 1}{\omega_q^2}\left[e^{-i\omega_q t} -1\right] + \frac{it}{\omega_q}\right)\,.\label{Kt}
%\label{cum2}
\end{align}
Converting the summation over $q$ to an integration $\sum_q \rightarrow \frac{\Vol}{(2\pi)^3 v_s^3} \int d^3 \omega$ (where $\Vol$ is the sample volume) and noting that $|\lambda_q|^2$ may be expressed in terms of the spectral density function $J(\omega)$ (see \Eq{Jw} in \App{Sec:Matrix_elements} below), we re-write \Eq{Kt} as
\begin{align}
K(t) &= \int_0^{\infty} d\omega \, J(\omega) \left(\frac{N_q}{\omega^2}\left[e^{i\omega t} - 1\right]\right.\nonumber\\
&\hspace{20mm}\left. + \frac{N_q + 1}{\omega^2}\left[e^{-i\omega t} -1\right] + \frac{it}{\omega}\right). \label{eq:fullK}
\end{align}
In the long-time limit, \Eq{eq:fullK} simplifies to
\begin{equation}
K(t \rightarrow \infty) = - i\Omega_p t - S,
\label{Kasympt}
\end{equation}
with the polaron shift
\be
\Omega_p = - \int_0^{\infty} d\omega \, \frac{J(\omega)}{\omega}\label{eq:pol_shift}
\ee
and the Huang-Rhys factor
\begin{align}
S &= \int_0^{\infty} d\omega \, \frac{J(\omega)}{\omega^2} \left(2N_q  + 1\right)\nonumber\\
&= \int_0^{\infty} d\omega \, \frac{J(\omega)}{\omega^2} \coth{\left(\frac{\omega}{2 k_B T}\right)}\,.
\label{Huang}
\end{align}

\begin{figure}[htp]
%\hskip-3.5cm
\includegraphics[scale=0.35]{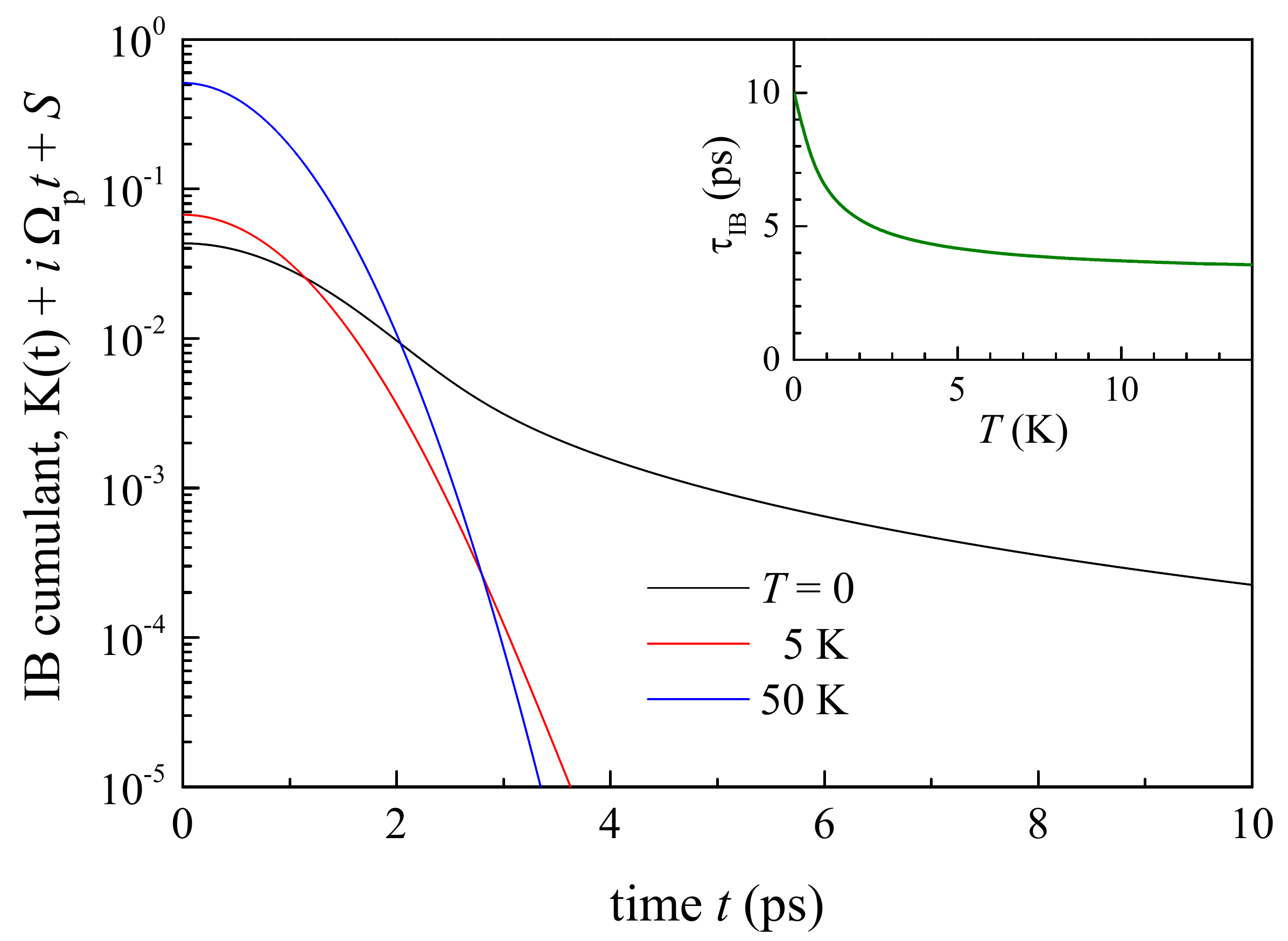}
%\vskip-2.3cm
\caption{IB model cumulant $K(t)$, with its long-time asymptotics $- i\Omega_p t - S$ subtracted, as a function of time $t$ for different temperatures as given. The parameters used are listed at the end of \App{Sec:Matrix_elements}. Inset: the phonon memory time $\tau_{\rm IB}$ playing the role of the cut-off parameter in calculation of the cumulants for different realizations in the $L$N approach.
}
\label{fig:Cumulant}
\end{figure}

Figure~\ref{fig:Cumulant} shows the cumulant function $K(t)$ of the IB model with the asymptotic behavior $-i\Omega_p t - S$ subtracted. The polaron timescale $\tau_{\rm IB}$ is the time taken for the remaining part of the cumulant, $K(t)+ i\Omega_p t + S$, to drop below a certain threshold value. 
The choice of this threshold is dictated by the accuracy required in the calculation: $\tau_{\rm IB}$ determines the choice of the minimal time step in the NN approximation ($\Delta t\approx \tau_{\rm IB}$) and the $L$N approach ($L \Delta t\approx \tau_{\rm IB}$), and any contributions from the quickly decaying part of the cumulant $K(t)+ i\Omega_p t + S$ beyond $t = \tau_{\rm IB}$ are neglected in the calculation. Choosing a threshold of $10^{-4}$, we see from \Fig{fig:Cumulant} that the polaron timescale $\tau_{\rm IB}$ is approximately 3.25\,ps at $T=5$ and $T=50$ K for the realistic QD parameters used in the calculation (see \App{Sec:Matrix_elements}). This timescale is, however, strongly dependent on the exciton confinement length $l$ and speed of sound in the material $v_s$ (set to 3.3 nm and $4.6 \times 10^3$\,m/s respectively to produce \Fig{fig:Cumulant}). We therefore define, in \Eq{eq:tauIB} of the main text, $\tau_{\rm IB}$ in terms of these key parameters.

At very low temperatures, $\tau_{\rm IB}$ also becomes temperature-dependent, as it is clear from \Fig{fig:Cumulant}; in the present case $\tau_{\rm IB}$ increases to 10\,ps at $T=0$. The full temperature dependence of $\tau_{\rm IB}$ is shown up to $T=14$ K in the inset of \Fig{fig:Cumulant}.

\section{Exciton-phonon coupling matrix element $\lambda_q$ and the spectral density function $J(\omega)$}
\label{Sec:Matrix_elements}

At low temperatures, the exciton-phonon interaction is dominated by the deformation potential coupling to longitudinal acoustic phonons. Assuming (i) that the phonon parameters in the confined QD do not differ significantly from those in the surrounding material, and (ii) that the acoustic phonons have linear dispersion $\omega_q = v_s|q|$, where $v_s$ is the sound velocity in the material, the matrix coupling element $\lambda_q$ is given by
\begin{equation}
\lambda_q = \frac{q \mathcal{D}(q)}{\sqrt{2 \rho_m \omega_q \Vol}}\,,
\end{equation}
where $\rho_m$ is the mass density of the material. Assuming a factorizable form of the exciton wave function, $\Psi_X(\mathbf{r}_e,\mathbf{r}_h)=\psi_e(\mathbf{r}_e)\psi_h(\mathbf{r}_h)$, where $\psi_{e(h)}(\mathbf{r})$ is the confined electron (hole) ground state wave function, the form-factor $\mathcal{D}(q)$ is given by
\begin{equation}
\mathcal{D}(q) = \int d\mathbf{r} \left[D_v|\psi_h(\mathbf{r})|^2 - D_c|\psi_e(\mathbf{r})|^2\right]e^{-i\mathbf{q}\cdot\mathbf{r}},
\end{equation}
with $D_{c(v)}$ being the material-dependent deformation potential constant for the conduction (valence) band.
%Details of the wavefunction do not significantly affect the optical linear response of the system.
We choose for simplicity spherically symmetric parabolic confinement potentials which give Gaussian ground state wave functions:
\begin{equation}
\psi_{e(h)}(\mathbf{r}) = \frac{1}{(\sqrt{\pi} l_{e(h)} )^{3/2}} \exp(-\frac{r^2}{2l_{e(h)}^2}),
\end{equation}
and thus
\begin{equation}
\lambda_q = \sqrt{\frac{q}{2\rho_m v_s \Vol}}(D_v - D_c) \,e^{-\frac{l^2q^2}{4}}\,,
\label{eq:Mq}
\end{equation}
taking the case of $l_e=l_h=l$ for simplicity.\\
\\
The spectral density $J(\omega)$  is defined as
\begin{equation}
J(\omega) = \sum_{q}|\lambda_q|^2 \delta(\omega - \omega_q).\label{eq:specdensity}
\end{equation}
This is equivalent to taking the product of $|\lambda_q|^2 $ with the density of states in $\omega$-space.
Switching from the summation to an integration, as in \Eq{Kt}, the spectral density becomes
\be
J(\omega)= |\lambda_q|^2 \frac{2 \Vol}{(2\pi)^2 v_s^3} \omega^2 =\frac{\omega^3 (D_c - D_v)^2}{4 \pi^2 \rho_m v_s^5}e^{-\frac{\omega^2}{\omega_0^2}}\, ,
\label{Jw}
\ee
where $q=\omega/v_s$ and $\omega_0 = \sqrt{2} v_s/l$ is the so-called ``cut-off'' frequency; it is inversely related to the phonon memory time, $\tau_{\text{IB}} \approx 2\pi/\omega_0$, leading to \Eq{eq:tauIB}.\\

In all calculations, we use $l = 3.3$\,nm, $D_c - D_v = -6.5$\,eV, $v_s = 4.6 \times 10^3$\,m/s, and $\rho_m = 5.65$\,g/cm$^3$.

\section{Long-time analytics for the linear polarization}
\label{Sec:long}

In this section, we derive the approximate analytic result Eqs.\,(\ref{eq:Panalyt}) and (\ref{eq:Htilde}) for the linear polarization $\hat{P}(t)$ in the long-time limit. This approximation is valid for small values of the exciton-cavity coupling strength $g$, which guarantees that the polariton timescale is much longer than the phonon memory time, $\tau_{\rm JC}\gg\tau_{\rm IB}$. As a starting point, we take the result for $\hat{P}(t)$ in the NN approach, Eqs.\,(\ref{eq:PNN}) and (\ref{eq:Gmatrix}), and use it for $\Delta t\gtrsim  \tau_{\rm IB}$. This condition implies that we can take both $K_0$ and $K_1$ in the long-time limit, using the asymptotic formula \Eq{eq:Kinf}:
\begin{align}
K_0 &= K(\Delta t) \approx - i\Omega_p\Delta t -S,
\label{K0}
\\
K_1 &= \frac{1}{2}\left(K(2\Delta t) - 2K(\Delta t)\right) \approx \frac{S}{2}\,.
\label{K1}
\end{align}
We would now like to replace the product of $N$ matrices in \Eq{eq:PNN} by an approximate analytic expression, taking the Trotter limit $N\to\infty$.
To do so, we initially derive explicit expressions for $\hat{M}$ and $\hat{G}$ in the two-state basis of $\ket{X}$ and $\ket{C}$. From Eq. (\ref{eq:M})  we obtain
\begin{equation}
%\hat{M}(\Delta t)
\begin{pmatrix}M_{XX} & M_{XC}\\M_{CX} & M_{CC} \end{pmatrix}
= e^{-i\omega_1 \Delta t}\begin{pmatrix}
1-\beta^2 \delta & -\alpha\beta\delta\\
-\alpha\beta\delta & 1-\alpha^2\delta
\end{pmatrix}\label{eq:gmat1},
\end{equation}
where $\omega_{1,2}$ are the eigenvalues of the Jaynes-Cummings Hamiltonian $H_{\rm JC}$, $\delta = 1 - e^{-i(\omega_2 - \omega_1)\Delta t}$, and $\alpha$ and $\beta$ make up the unitary matrices $\hat{Y}$, $\hat{Y}^{-1}$ that diagonalize $H_{\rm JC}$:
\begin{align}
H_{\text{JC}} & =\begin{pmatrix}
\omega_X & g\\
g & \omega_C
\end{pmatrix}=
\hat{Y}^{-1} \begin{pmatrix}
\omega_1 & 0\\
0 & \omega_2
\end{pmatrix} \hat{Y}\,,
\label{eq:eigenmatrix}\\
\hat{Y} &= \begin{pmatrix}
\alpha & -\beta\\
\beta & \alpha
\end{pmatrix}\,,\label{eq:Ytransform}\\
\alpha &= \frac{\Delta}{\sqrtsign{\Delta^2 + g^2}},\\
\beta &= \frac{g}{\sqrtsign{\Delta^2 + g^2}},\\
\omega_{1,2} &= \frac{\omega_X + \omega_C}{2} \pm \sqrtsign{g^2 +\delta^2},
\end{align}
with $\Delta = \sqrtsign{\delta^2 + g^2} - \delta$ and $\delta = \nicefrac{1}{2}\left(\omega_X - \omega_C\right)$. Substituting the expression for $\hat{M}$ given by \Eq{eq:gmat1} into \Eq{eq:Gmatrix}, and using Eqs.\,(\ref{K0}) and (\ref{K1}), we find
\begin{align}
\hat{G} &=
\begin{pmatrix}
 M_{XX}e^{K_0+2K_1} & M_{XC}\\ M_{CX}e^{K_0} & M_{CC} \label{eq:Gmatrix1}
 \end{pmatrix}\nonumber\\
&\approx e^{-i\omega_1 \Delta t} \begin{pmatrix}
e^{-i\Omega_p\Delta t} (1-\beta^2\delta) &-\alpha\beta\delta\\
-e^{-i\Omega_p\Delta t-S}\alpha\beta\delta &1-\alpha^2\delta
\end{pmatrix}.
\end{align}

Now we use the fact that $\Delta t \ll \tau_{\rm JC}$ (which is equivalent to $|\omega_2 - \omega_1|\Delta t \ll 1$). We also assume that the polaron shift $\Omega_p$ is small, so that $|\Omega_p|\Delta t \ll 1$. Working within these limits is equivalent to taking the Trotter limit $\Delta t=t/N\to 0$. Keeping only the terms linear in $\Delta t$ in the matrix elements, we obtain
\begin{equation}
\hat{G} \approx e^{-i\omega_1 \Delta t} \left[ \mathbb{1} - i\Delta t\begin{pmatrix}
\Omega_p + \beta^2 \omega_{21} & \alpha\beta\omega_{21}\\
\alpha\beta\omega_{21}e^{-S} & \alpha^2\omega_{21}
\end{pmatrix}\right],
\label{eq:gmat2}
\end{equation}
where $\omega_{21} = \omega_2 - \omega_1$ and  $\mathbb{1}$ is a $2\times2 $ identity matrix. From Eq. (\ref{eq:eigenmatrix}) and the fact that $\alpha^2+\beta^2=1$ we find
\begin{align*}
\beta^2(\omega_2 - \omega_1) &= \omega_X-\omega_1,\\
\alpha^2(\omega_2-\omega_1) &= \omega_C-\omega_1,\\
\alpha\beta(\omega_2-\omega_1) &=g.
\end{align*}
This allows us to re-write \Eq{eq:gmat2} in the following way
\begin{equation*}
\hat{G} = e^{-i\omega_1 \Delta t} \left[\mathbb{1}(1+i\omega_1 \Delta t) - i\Delta t\begin{pmatrix}
\omega_X + \Omega_p & g\\
ge^{-S} & \omega_C
\end{pmatrix}\right].
\end{equation*}
Now, we diagonalize $\hat{G}$:
\be
\hat{G} = \hat{Z}\hat{\Lambda}\hat{Z}^{-1}\,,
\label{eq:gylambday}
\ee
where the transformation matrix has the form
\be
\hat{Z} = \begin{pmatrix}
e^{S/2} & 0\\
0 & 1
\end{pmatrix}
\begin{pmatrix}
\tilde{\alpha} & \tilde{\beta}\\
-\tilde{\beta} & \tilde{\alpha}
\end{pmatrix},
\ee
in which the second matrix diagonalizes a phonon-renormalized JC Hamiltonian $\tilde{H}$, as defined in \Eq{eq:Htilde},
\begin{align}
\tilde{H}&=
\begin{pmatrix}
\omega_X + \Omega_p & ge^{-S/2}\\
ge^{-S/2} & \omega_C
\end{pmatrix} \nonumber\\
&= \begin{pmatrix}
\tilde{\alpha} & \tilde{\beta}\\
-\tilde{\beta} & \tilde{\alpha}
\end{pmatrix}
\begin{pmatrix}
\tilde{\omega}_1 & 0\\
0 & \tilde{\omega}_2
\end{pmatrix}
\begin{pmatrix}
\tilde{\alpha} & -\tilde{\beta}\\
\tilde{\beta} & \tilde{\alpha}
\end{pmatrix}.
\end{align}
The matrix of the eigenvalues $\hat{\Lambda}$ in \Eq{eq:gylambday} then takes the form
\begin{equation}
\hat{\Lambda} = e^{-i\omega_1 \Delta t}\left[\mathbb{1} - i\Delta t \begin{pmatrix}
\tilde{\omega}_1 - \omega_1 & 0\\
0 & \tilde{\omega}_2-\omega_1
\end{pmatrix}\right].
\end{equation}
Coming back to the NN expression for the polarization \Eq{eq:PNN},
\begin{equation}
\hat{P}(t) = \begin{pmatrix}
 e^{K_0} & 0\\ 0 & 1
\end{pmatrix}
\hat{G}^{N}\hat{G}^{-1}\hat{M},
\end{equation}
we note that $\hat{G}^{-1} \approx \mathbb{1}$ and $\hat{M} \approx \mathbb{1}$ in the limit $\Delta t\to 0$,
and also $e^{K_0} \approx e^{-S}$ (still keeping the condition $\Delta t\gtrsim \tau_{\rm IB}$). We then obtain in the long-time limit $t\gtrsim \tau_{\text{IB}}$:
\begin{equation}
\hat{P}(t) = e^{-i\omega_1 t}\begin{pmatrix}
e^{-S} & 0\\
0 & 1
\end{pmatrix} \hat{Z} \hat{\Lambda}^N \hat{Z}^{-1}\,.\label{eq:P_t_analytic_lim}
\end{equation}

Finally, we take the limit $N\to \infty$ in the expression $\hat{\Lambda}^N$, using an algebraic formula
$$
\lim_{N \to \infty} \left(1 + \frac{x}{N}\right)^N = e^x\,.
$$
Introducing
\begin{align*}
x &= -i(\tilde{\omega}_1 - \omega_1)t\,,\\
y &= -i(\tilde{\omega}_2 - \omega_1)t\,,
\end{align*}
we find
\begin{align}
\lim_{N \to \infty}\hat{\Lambda}^N &= \lim_{N \to \infty} \begin{pmatrix}
1 + \frac{x}{N} & 0\\
0 & 1 + \frac{y}{N}
\end{pmatrix}^N
\nonumber\\
&= e^{i\omega_1 t} \begin{pmatrix}
e^{-i\tilde{\omega}_1 t} & 0\\
0 & e^{-i\tilde{\omega}_2 t}
\end{pmatrix}.\label{eq:lambdaN}
\end{align}
Substituting \Eq{eq:lambdaN} into \Eq{eq:P_t_analytic_lim} we arrive at \Eq{eq:Panalyt} of the main text.

\section{Refined full time analytic approach}
\label{Sec:refined}
The analytic solution derived in \App{Sec:long} is suited only for describing the optical polarization at long times $t\gtrsim  \tau_{\rm IB}$, so that any information on the evolution at short times, which is responsible for the so-called phonon broadband observed in the optical spectra of quantum dots, is missing. To improve on this, we derive a refined, purely analytic approach which properly takes into account both the short and long time dynamics, providing  a smooth transition between the two regimes.

We again start with the general formula \Eq{eq:Pdef} for the linear polarization, writing it in a matrix form using the two basis states $\ket{X}$ and $\ket{C}$:
\be
\hat{P}(t)= \langle \hat{U}(t)\rangle\,.
\label{pt}
\ee
Note that the expectation value  in \Eq{pt} is taken over the phonon system in thermal equilibrium, and the $2\times 2$ evolution  matrix operator $\hat{U}(t)$ has the form:
\be
\hat{U}(t)= e^{iH_{\rm ph}t}e^{-iHt}=e^{-iH_{\rm JC}t}e^{iH_1t}e^{-iHt}\,,\label{eq:ev_op1}
\ee
where
\begin{align}
H_1 &= H_{\rm JC}+ H_{\rm ph}\mathbb{1},\\
H &= H_1+
\begin{pmatrix}1 &  0\\ 0& 0\end{pmatrix} V\,,
\end{align}
with $H_{\rm JC}$ ($H_{\rm ph}$ and $V$) defined in \Eq{eq:HJC} (\Eq{eq:HIB}) of the main text. We apply the polariton transformation, defined in \Eq{eq:eigenmatrix}, to \Eq{eq:ev_op1} for the evolution operator $\hat{U}(t)$, 
\begin{equation}
\hat{U}(t)=\hat{Y}^{-1} e^{-iH_0t}e^{i\bar{H}_1t}e^{-i\bar{H}t} \hat{Y}\,,\label{eq:ev_op2}
\end{equation}
where $H_0$ is a $2\,\times\,2$ matrix of eigenvalues of $H_{\rm JC}$,
\begin{equation}
H_0 = \begin{pmatrix} \omega_1 & 0\\0 &\omega_2 \end{pmatrix}\,,
\end{equation}
and
\bea
\bar{H}_1&=&\hat{Y} H_1 \hat{Y}^{-1}= H_0+ H_{\rm ph}\mathbb{1}\,,\\
\bar{H}&=&\hat{Y} H \hat{Y}^{-1}= H_0+ H_{\rm ph}\mathbb{1}+\hat{Q} V\,,\\
\hat{Q}&=&\hat{Y} \begin{pmatrix} 1 & 0 \\ 0 & 0 \end{pmatrix} \hat{Y}^{-1} =\begin{pmatrix} \alpha^2 & \alpha\beta \\ \alpha\beta & \beta^2 \end{pmatrix}.
\eea
We now define a reduced evolution operator, $\bar{U}(t)$, such that \Eq{eq:ev_op2} may be re-expressed as
\begin{equation}
\hat{U}(t) = \hat{Y}^{-1} e^{-iH_0t} \bar{U}(t) \hat{Y}\,.
\end{equation}
Expressing $\bar{U}(t)$ as an exponential series,
\be
\bar{U}(t)=  e^{i\bar{H}_1t}e^{-i\bar{H}t}= \mathcal{T} \exp\left\{-i\int_0^t H_{int}(t') dt'\right\}\,,
\ee
where
\be
H_{int}(t) = e^{i\bar{H}_1t} (\bar{H}-\bar{H}_1) e^{-i\bar{H}_1t}= \hat{Q}(t) V(t)\,,
\ee
with individual interaction representations of the polariton and phonon operators: $\hat{Q}(t)= e^{iH_0t} \hat{Q}e^{-iH_0t}$ and $V(t)=e^{iH_{\rm ph}t} V e^{-iH_{\rm ph}t}$. The expectation value of $\bar{U}(t)$ then becomes an infinite perturbation series:
\begin{align}
\langle \bar{U}(t)\rangle &= \mathbb{1}+(-i)^2 \int _0^t dt_1  \int _0^{t_1} dt_2 \hat{Q}(t_1)\hat{Q}(t_2) \langle V(t_1)V(t_2)\rangle \nonumber\\
& \hspace{5mm} + \cdots
\label{Ubar}
\end{align}
Using Wick's theorem, all of the expectation values split into pair products. For example,
\begin{align*}
\langle V(t_1)&V(t_2)V(t_3)V(t_4)\rangle = D(t_1-t_2)D(t_3-t_4)\\
&+D(t_1-t_3)D(t_2-t_4)+D(t_1-t_4)D(t_2-t_3)\,,
\end{align*}
where
\begin{equation*}
D(t-t')=\langle V(t)V(t')\rangle = \sum_q|\lambda_q|^2 i D_q(t-t')
\end{equation*}
is the full phonon propagator, see \Eq{Ddef}.

\begin{figure}[htp]
\begin{tikzpicture}[baseline=(current bounding box.north)]

   \newcommand{\midarrow}{\tikz \draw[-{Straight Barb[angle=60:.2cm 1]},line width=.03cm] (0,0) -- +(.1,0);}

  \node [font=\huge] at (0,0.5) {$\hat{\Sigma}$};
  \node at (0.65,0.4) {$=$};
  \node at (3.5,0.4) {$+$};
  \node at (0.65,-2) {$+$};
  \node at (6.3,-2) {$+ \, \cdots$};

  %first diagram
  % Define coordinates
  \def\Radius{1}
  \path
    (2-\Radius, 0) coordinate (A1)
    -- coordinate (M1)
    (2+\Radius, 0) coordinate (B1)
  ;
  % Draw semicircle
  \draw[dashed]
    (B1) arc (0:180:\Radius);
    \draw (A1)-- node {\midarrow} (B1);
  ;
  % Annotations
  \path[inner sep=0pt]
    (A1) node[below=.3333em] {$t$}
    (B1) node[below=.3333em] {$t'$}
  ;

  %second diagram
  % Define coordinates
  \path
    (5-\Radius, 0) coordinate (A2)
    -- coordinate (M2)
    (5+\Radius, 0) coordinate (B2)
    (5,0) coordinate (C2)
    (7,0) coordinate (D2)
  ;
  % Draw semicircle
  \draw[dashed]
    (B2) arc (0:180:\Radius);
  \draw[dashed]
    (D2) arc (0:180:\Radius);
  \draw (A2)-- node {\midarrow} (C2);
  \draw (C2)-- node {\midarrow} (B2);
  \draw (B2)-- node {\midarrow} (D2);
  ;
  % Annotations
  \path[inner sep=0pt]
    (A2) node[below=.3333em] {$t$}
    (B2) node[below=.3333em] {$t_2$}
    (C2) node[below=.3333em] {$t_1$}
    (D2) node[below=.3333em] {$t'$}
  ;

  %third diagram
  % Define coordinates
  \path
    (1.25, -3) coordinate (A2)
    (2.25, -3) coordinate (B2)
    (4.25,-3) coordinate (C2)
    (5.25,-3) coordinate (D2)
  ;
  % Draw semicircle
  \draw[dashed]
    (C2) arc (0:180:\Radius);
  \draw[dashed]
    (D2) arc (0:180:2);
  \draw (B2)-- node {\midarrow} (C2);
  \draw (A2)-- node {\midarrow} (B2);
  \draw (C2)-- node {\midarrow} (D2);
  ;
  % Annotations
  \path[inner sep=0pt]
    (A2) node[below=.3333em] {$t$}
    (B2) node[below=.3333em] {$t_1$}
    (C2) node[below=.3333em] {$t_2$}
    (D2) node[below=.3333em] {$t'$}
  ;

\end{tikzpicture}
\caption{Second and fourth order diagrams contributing to the full self energy. Solid lines with arrows (dashes lines) represent the polariton (phonon) non-interacting Green's functions.}
\label{Fig:diagr}
\end{figure}
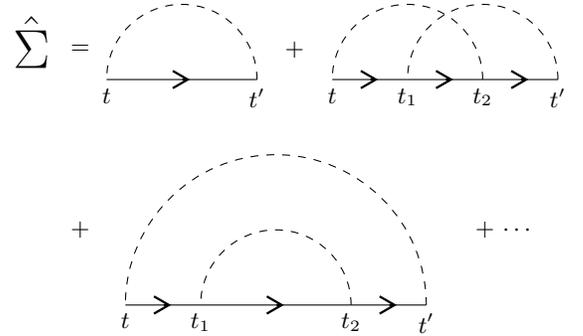

It is convenient to introduce the bare polariton Green's function
\begin{align}
\hat{G}^{(0)}(t)&=\begin{pmatrix} G_1^{(0)}(t) & 0 \\ 0 & G_2^{(0)}(t) \end{pmatrix}\nonumber\\
&=\theta(t)\begin{pmatrix} e^{-i\omega_1t} & 0 \\ 0  & e^{-i\omega_2t} \end{pmatrix},
\label{G0}
\end{align}
where $\theta(t)$ is the Heaviside step function. Then the full phonon-dressed polariton Green's function $\hat{G}(t)$, which is related to the polarization matrix via
\be
\hat{P}(t)=\hat{Y}^{-1}\hat{G}(t)\hat{Y}\,,
\label{PY}
\ee
satisfies the following Dyson's equation:
\begin{align}
\hat{G}(t)&=\hat{G}^{(0)}(t)\nonumber\\
&+ \int_{-\infty}^\infty dt_1 \int_{-\infty}^\infty dt_2  \, \hat{G}^{(0)}(t-t_1) \hat{\Sigma}(t_1-t_2) \hat{G}(t_2)\,.
\label{Gtfull}
\end{align}
Note that this equation is equivalent to the perturbation series \Eq{Ubar}. Here, the self energy $\hat{\Sigma}$ is represented  by all possible connected diagrams such as the 2nd and 4th order diagrams sketched in \Fig{Fig:diagr}, which are given by the following expressions:
\begin{align}
&\hat{\Sigma}(t-t')= \hat{Q}\hat{G}^{(0)}(t-t')\hat{Q} D(t-t')\nonumber\\
&\hspace{5mm}+ \int_{-\infty}^\infty dt_1 \int_{-\infty}^\infty dt_2 \, \Big\{\hat{Q}\hat{G}^{(0)}(t-t_1)\hat{Q} 
\nonumber\\
&\hspace{7mm} \times \hat{G}^{(0)}(t_1-t_2) \hat{Q} \hat{G}^{(0)}(t_2-t')\hat{Q}\nonumber\\
&\hspace{7mm} \times [D(t-t_2)D(t_1-t')+D(t-t')D(t_1-t_2)]\Big\}\nonumber\\
&\hspace{5mm}+\dots \,.
\label{Sigma}
\end{align}

\begin{figure}[htp]
%\hskip-40mm
\includegraphics[scale=0.35]{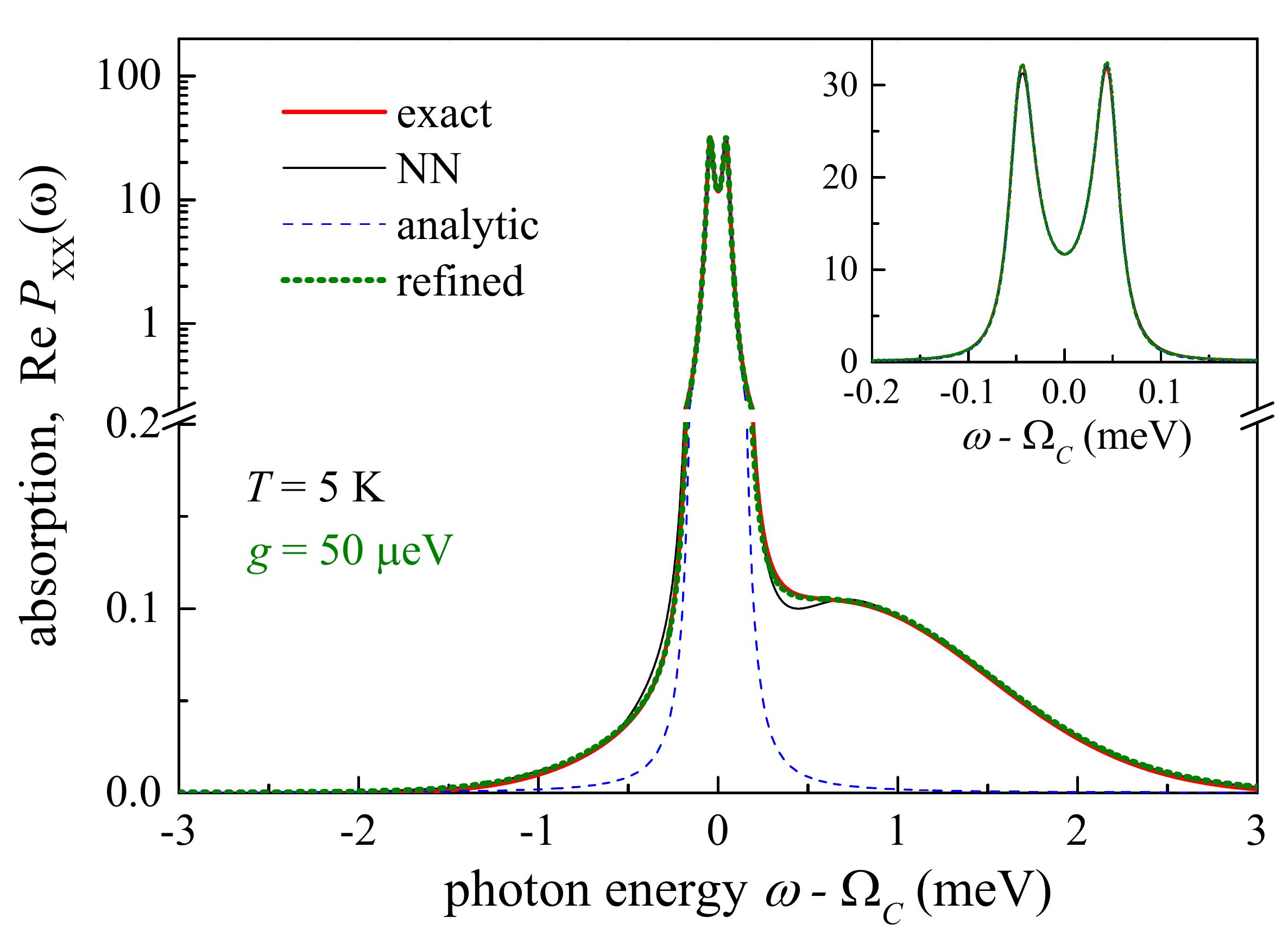}
%\vskip-25mm
\caption{Absorption spectra for $g=50\,\mu$eV, $T=5\,$K, and zero detuning, calculated in the $L$N approach with $L=15$ (red thick solid lines), NN approach with $L=1$ (black thin solid lines), long-time analytic approximation (blue dashed lines) and refined analytics (green dotted line). Other parameters used: $\Omega_X=1329.6$\,meV, $\gamma_X = 2\,\mu$eV, $\Omega_C=\Omega_X + \Omega_p$ with $\Omega_p = -49.8\,\mu$eV, and $\gamma_C = 30\,\mu$eV.}
\label{Fig:refined}
\end{figure}
Equations (\ref{Gtfull}) and (\ref{Sigma}) are {\it exact} provided that all the connected diagrams are included in the self energy. No approximations have been used so far.

\begin{figure}[htp]
%\hspace{-3.5mm}
%\begin{subfloat}[t]{\columnwidth}
%\vskip 5pt
\includegraphics[scale=0.41]{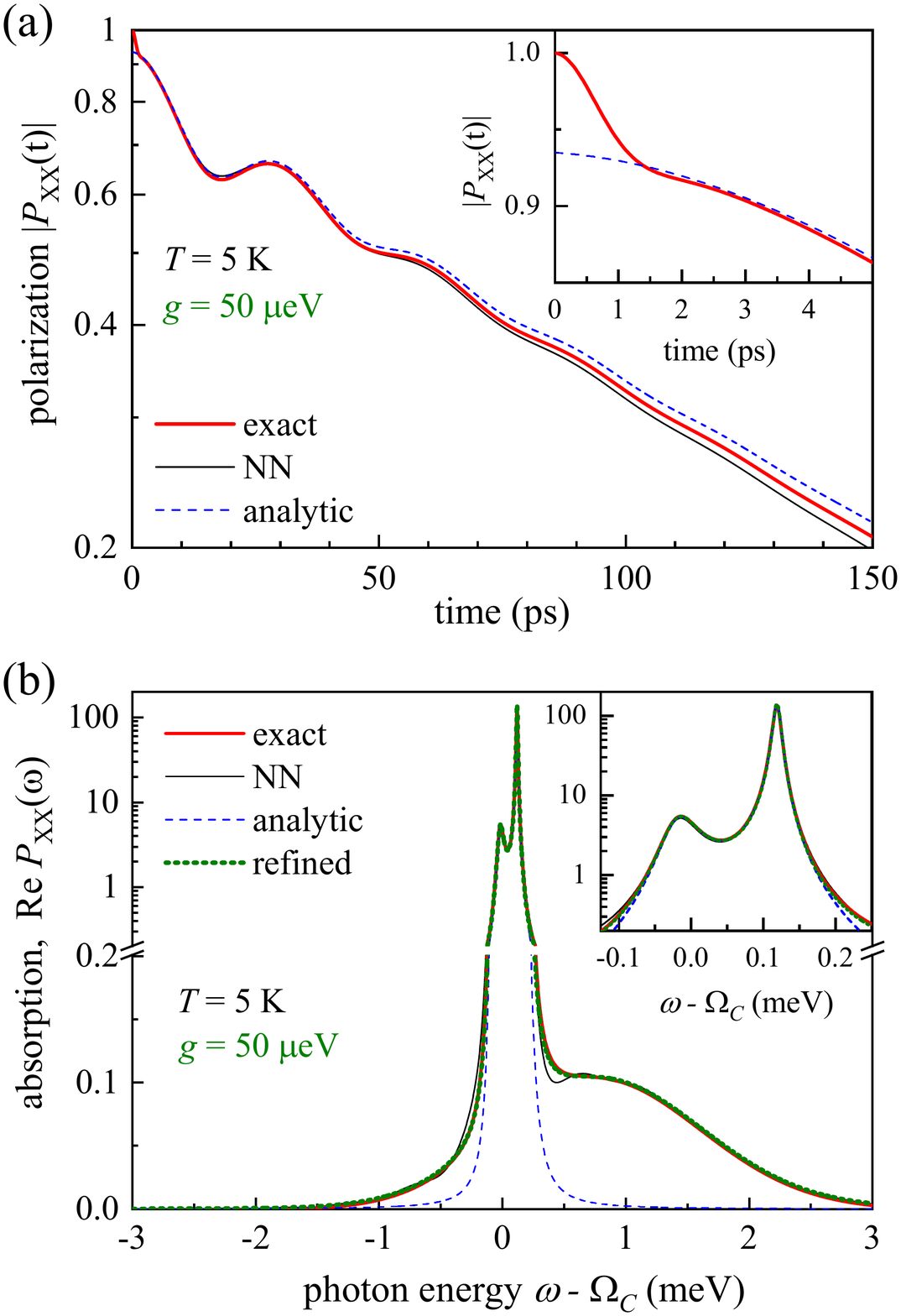}
%\hskip20mm
%\end{subfloat}

%\hspace{-2mm}
%\begin{subfloat}[t]{\columnwidth}
%\vskip 0pt
%\includegraphics[scale=0.35]{Det5Kb}
%\end{subfloat}
%\vskip-25mm
\caption{(a) Excitonic linear polarization and (b) absorption for  $g=50\,\mu$eV, $T=5$\,K, and nonzero detuning, calculated in the $L$N approach with $L=15$ (red thick solid lines), NN approach with $L=1$ (black thin solid lines), analytic approximation (blue dashed lines) and refined analytics (green dotted line). Other parameters used: $\Omega_X=1329.6$\,meV, $\gamma_X = 2\,\mu$eV, $\Omega_C=1329.45$\,meV, and $\gamma_C = 30\,\mu$eV.
}
\label{fig:Det1}
\end{figure}

\begin{figure}[htp]
%\hspace{-4.5mm}
%\begin{subfigure}[t]{\columnwidth}
%\vskip 5pt
%\hskip-35mm
\includegraphics[scale=0.41]{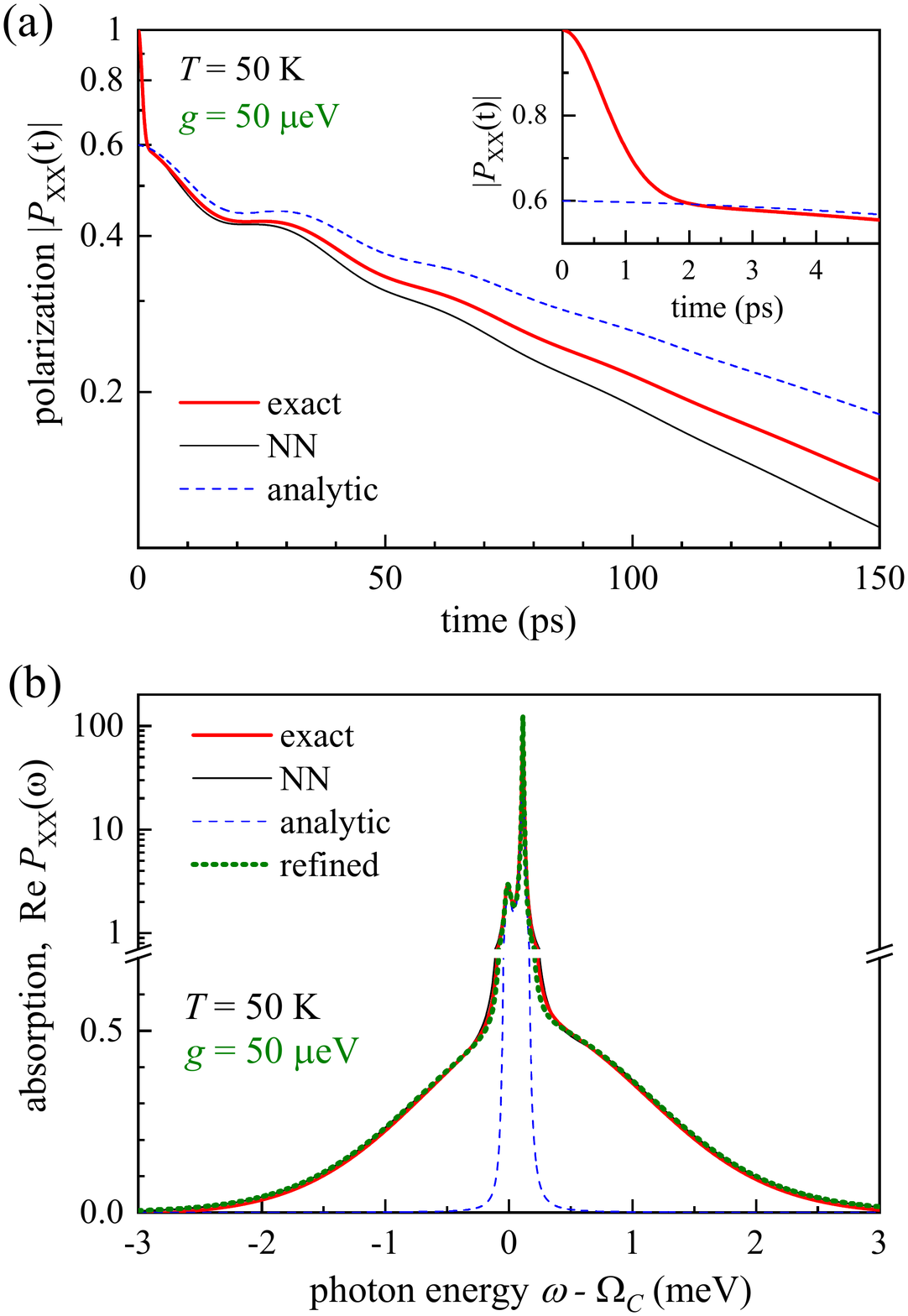}
%\end{subfigure}
%\hskip1mm
%\begin{subfigure}[t]{\columnwidth}
%\vskip 0pt
%\hskip15mm
%\includegraphics[scale=0.35]{Det50Kb}
%\vskip8mm
%\end{subfigure}
%\vskip-25mm
\caption{As \Fig{fig:Det1} but for $T=50$\,K.}
\label{fig:Det2}
\end{figure}

In the case of isolated (phonon-decoupled) polariton states, all of the matrices are diagonal and the problem reduces to the IB model for each polariton level, having an exact analytic solution which we exploit in our approximation. For the two phonon-coupled polariton states treated here, the exact solvability is hindered by the fact that the matrices $\hat{Q}$ and $\hat{G}^{(0)}(t)$ do not commute for any finite time $t$. However, in the timescale $|\omega_1-\omega_2|t\ll 1$, \Eq{G0} may be approximated as $\hat{G}^{(0)}(t)\approx \theta (t) e^{-i\omega_1t} \mathbb{1}$ and thus $\hat{G}^{(0)}(t)$ approximately commutes with $\hat{Q}$, so for example,
\begin{align*}
 \hat{Q}\hat{G}^{(0)}&(t-t_1)\hat{Q} \hat{G}^{(0)}(t_1-t_2)\hat{Q} \hat{G}^{(0)}(t_2-t')\hat{Q}  \\
 &\approx \hat{Q} \hat{G}^{(0)}(t-t')\theta(t-t_1)\theta(t_1-t_2)\theta(t_2-t')\,,
\end{align*}
using $\hat{Q} ^2=\hat{Q}$. Clearly, this approximation is valid if $\tau_{\rm JC}\gg \tau_{\rm IB}$. In this case we obtain
\be
\hat{\Sigma}(t)=\hat{Q}\begin{pmatrix}\Sigma_1(t) & 0 \\ 0 & \Sigma_2(t) \end{pmatrix},
\label{SE}
\ee
where $\Sigma_j(t)$ is the self energy of an isolated polariton state $j$, which contributes to the corresponding IB model problem
\begin{align}
{G}^{\rm IB}_j(t)&=G_j^{(0)}(t)\nonumber\\
&+\int_{-\infty}^\infty dt_1 \int_{-\infty}^\infty dt_2  G_j^{(0)}(t-t_1) \Sigma_j(t_1-t_2) {G}^{\rm IB}_j(t_2)\,,
\label{GIB}
\end{align}
having the following exact solution:
\be
{G}^{\rm IB}_j(t)=G_j^{(0)}(t)e^{K(t)}\,,
\label{GIBcum}
\ee
where the cumulant $K(t)$ is given by \Eq{Cum}. Equation~(\ref{GIB}) then allows us to find the self energies in frequency domain:
\be
\Sigma_j(\omega) = \frac{1}{G_j^{(0)}(\omega)}- \frac{1}{{G}^{\rm IB}_j(\omega)}\,,
\label{SEj}
\ee
where $\Sigma_j(\omega)$, $G_j^{(0)}(\omega)$, and $G^{\rm IB}_j(\omega)$ are the Fourier transforms of $\Sigma_j(t)$, $G_j^{(0)}(t)$, and $G^{\rm IB}_j(t)$, respectively.
The full matrix Green's function (and hence the polarization) is then obtained by solving Dyson's equation (\ref{Gtfull}) in frequency domain:
\be
\hat{G}(\omega)=\left[\mathbb{1}-\hat{G}^{(0)}(\omega)\hat{\Sigma}(\omega)\right]^{-1} \hat{G}^{(0)}(\omega)\,,
\label{Solution}
\ee
where $\hat{G}^{(0)}$ and $\hat{\Sigma}$ are given, respectively, by Eqs.\,(\ref{G0}) and (\ref{SE}), with self energy components provided via \Eq{SEj} by the IB model solution \Eq{GIBcum}.

An obvious drawback of the above analytic model is that it does not shows any phonon-induced renormalization of the exciton-cavity coupling due to the interaction with the phonon bath. This is a consequence of the present approach not properly taking into account the cumulative effect of self-energy diagrams of higher order, for which the approximate commutation of matrices $\hat{Q}$ and $\hat{G}^{(0)}(t)$ is not valid. But we know from the IB model that its exact solution in the form of a cumulant includes a nonvanishing contribution of all higher-order diagrams of the self energy series (for realistic phonon parameters of semiconductor quantum dots). This significant problem can, however, be easily healed through use of the large time asymptotics obtained in \App{Sec:long}. We introduce {\it by hand} one minor correction: we replace the exciton-cavity coupling $g$ in the bare JC Hamiltonian by the renormalized coupling strength $g e^{-S/2}$ in the following way
\be
H_{\rm JC}=
\begin{pmatrix} \omega_X & g \\ g & \omega_C \end{pmatrix}
\to\begin{pmatrix} \omega_X & ge^{-S} \\ g & \omega_C \end{pmatrix}. \label{modHJC}
\ee
As in \Eq{PY}, we can express the Fourier transform of the polarization as
\be
\hat{P}(\omega)= \begin{pmatrix} e^{-S/2} & 0 \\ 0 & 1 \end{pmatrix}
\begin{pmatrix}\bar{\alpha} & \bar{\beta} \\ -\bar{\beta} & \bar{\alpha}  \end{pmatrix}
\hat{\bar{G}}(\omega)
\begin{pmatrix}\bar{\alpha} & -\bar{\beta} \\ \bar{\beta} & \bar{\alpha}  \end{pmatrix}
\begin{pmatrix}e^{S/2} & 0 \\ 0 & 1 \end{pmatrix}, \label{Prev}
\ee
where the matrices containing $\bar{\alpha}$ and $\bar{\beta}$ diagonalize a symmetrized Hamiltonian $\bar{H}_{\rm JC}$:
\begin{align}
\bar{H}_{\rm JC} &= \begin{pmatrix} \omega_X & ge^{-S/2} \\ ge^{-S/2} & \omega_C \end{pmatrix}\nonumber\\
& =  \begin{pmatrix} \bar{\alpha} & \bar{\beta} \\ -\bar{\beta} & \bar{\alpha}  \end{pmatrix} \begin{pmatrix}\bar{\omega}_1 & 0 \\ 0 & \bar{\omega}_2 \end{pmatrix} \begin{pmatrix} \bar{\alpha} & -\bar{\beta} \\ \bar{\beta} & \bar{\alpha}  \end{pmatrix}.
\end{align}
Note that the first and last matrices of \Eq{Prev} arise as a result of the replacement of the adjusted Hamiltonian in \Eq{modHJC} with its symmetrized version $\bar{H}_{\rm JC}$. We see that $\hat{\bar{G}}(\omega)$ in \Eq{Prev} is the analog of \Eq{Solution} with a replacement $\alpha \rightarrow \bar{\alpha}$, $\beta \rightarrow \bar{\beta}$, $\omega_{1,2} \rightarrow \bar{\omega}_{1,2}$.

For $P_{XX}(\omega)$ and  $P_{CC}(\omega)$ the solution \Eq{Prev} gives the following simple explicit expressions:
\begin{align}
P_{XX}(\omega)&=\frac{\bar{\alpha}^2 \bar{G}_1^{(0)}(\omega)+ \bar{\beta}^2 \bar{G}_2^{(0)}(\omega)}{\bar{D}(\omega)}\,,\\
P_{CC}(\omega)&=\left( \frac{\bar{\alpha}^2}{\bar{G}_1^{\rm IB}(\omega)}+\frac{\bar{\beta}^2}{\bar{G}_2^{\rm IB}(\omega)}\right)\frac{\bar{G}_1^{(0)}(\omega)\bar{G}_2^{(0)}(\omega)}{\bar{D}(\omega)}\,,
\label{PXXCC}
\end{align}
where
\be
\bar{D}(\omega)=\bar{\alpha}^2\frac{\bar{G}_1^{(0)}(\omega)}{\bar{G}_1^{\rm IB}(\omega)}+\bar{\beta}^2\frac{\bar{G}_2^{(0)}(\omega)}{\bar{G}_2^{\rm IB}(\omega)}
\label{Den}
\ee
and $\bar{G}_j^{(0)}(\omega)$ and ${\bar{G}_j^{\rm IB}(\omega)}$ are, respectively, the Fourier transform of $\bar{G}_j^{(0)}(t)=\theta(t) e^{-i\bar{\omega}_j t}$ and ${\bar{G}_j^{\rm IB}(t)}=\bar{G}_j^{(0)}(t) e^{K(t)}$.

Figures~\ref{Fig:refined}, \ref{fig:Det1}(b) and \ref{fig:Det2}(b), as well as Figure \ref{fig:P_small_g}(b) of the main text, demonstrate a very good agreement between the refined analytic solution and the exact result provided by the full $L$N approach (with $L=15$). In addition to the case of zero detuning at low temperature ($T=5$\,K) presented in \Fig{Fig:refined}, we also show in Figs.\,\ref{fig:Det1} and \ref{fig:Det2} both low and high temperature results for a non-zero detuning of 0.1\,meV (the exact parameters are given in the captions).

\begin{figure}[h]
\centering
%\vskip35mm
%\hspace{-5mm}
%\begin{subfigure}[t]{\columnwidth}
%\vskip 5pt
%\hskip-33mm
\includegraphics[scale=0.42]{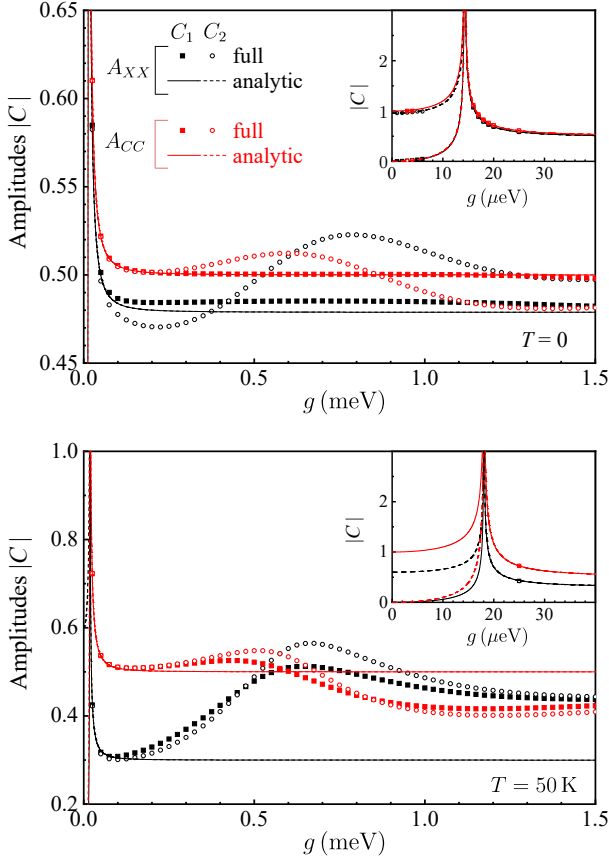}
%\end{subfigure}
%\hskip1mm
%\begin{subfigure}[t]{\columnwidth}
%\vskip 0pt
%\hskip15mm
%\includegraphics[scale=0.35]{Ampb}
%\vskip8mm
%\end{subfigure}
%\vskip-25mm
\caption{Polariton amplitude coefficient $|A_1|$ ($|A_2|$) as a function of the quantum dot-cavity coupling strength $g$ for (a) $T=0$ and (b) $T=50$\,K shown for the full calculation by full squares (open circles) and for the long-time analytic model by full (dashed) lines. Insets zoom in the region of small $g$, where the analytic model predicts significant changes of the amplitudes with $g$.}
\label{fig:Amp}
\end{figure}
\section{Polariton parameters and discussion of errors}
\label{sec:vary_g}

Having shown the behavior of the real polariton frequencies $\omega_{1,2}$ and linewidths $\Gamma_{1,2}$ in \Fig{fig:Gamma_vs_g} of the main text, we provide for completeness the amplitudes of the bi-exponential fit \Eq{eq:Pomega_pol} in \Fig{fig:Amp}. This figure addresses both excitonic and photonic polarization, $P_{XX}$ and $P_{CC}$ (black and red respectively), comparing results from the full calculation in the 15 neighbor approach (symbols) and the analytic approximation \Eq{eq:Panalyt} (lines).

\begin{figure}[htp]
\vspace{6mm}
\includegraphics[scale=0.35]{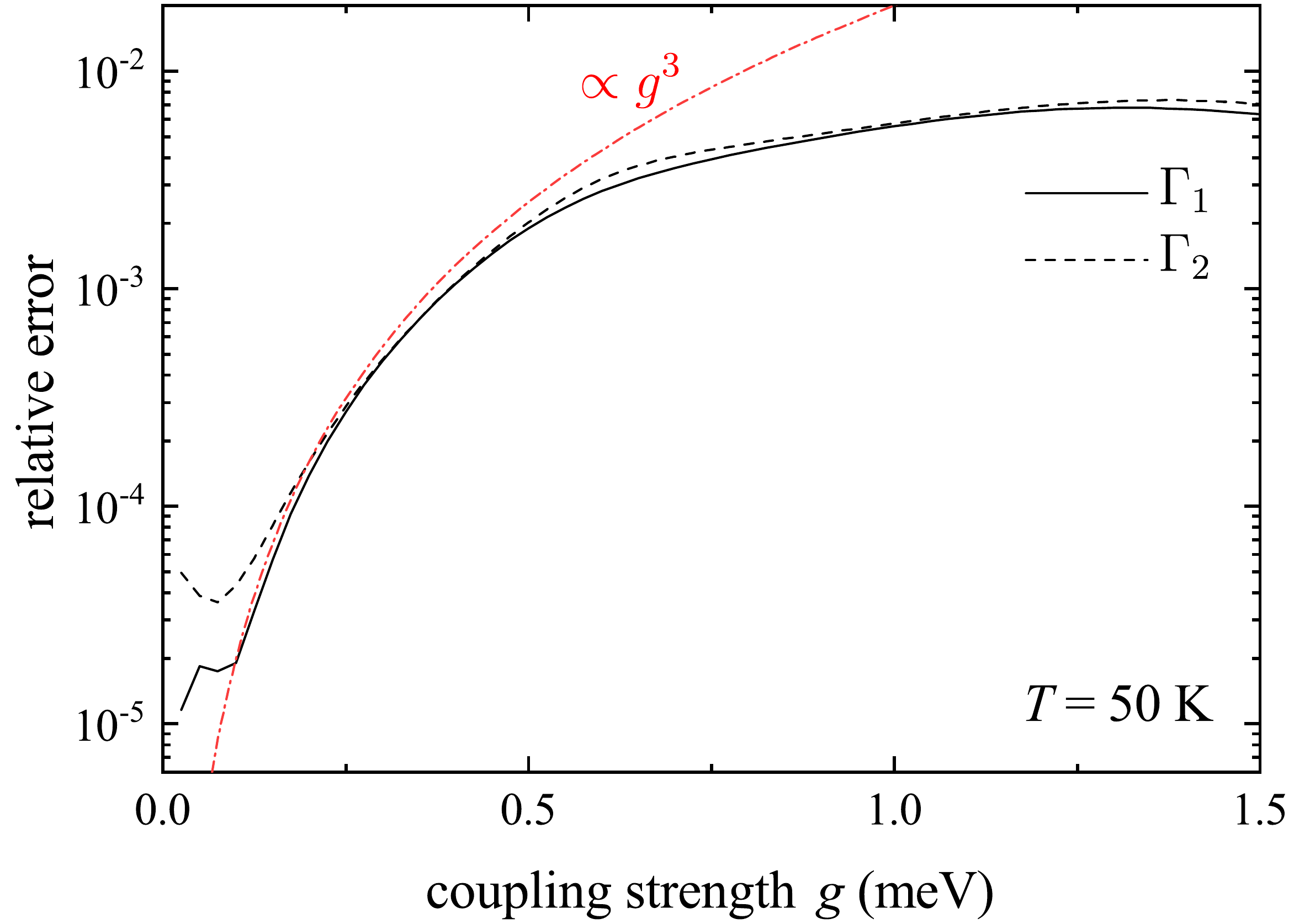}
\caption{Estimated relative error in polariton state linewidths $\Gamma_{1,2}$ at $T$=0 K and $T$=50 K, using the $L$N approach with $L = 13$, 14 and 15.}
\label{fig:Error_L15_calc}
\end{figure}
 
%\section{Estimation of error in polariton parameters}
Figure~\ref{fig:Error_L15_calc} shows the error in calculation of the linewidths $\Gamma_{1,2}$ via the Trotter decomposition as function of the coupling strength $g$. This error was estimated as the arithmetic average of the errors for $L=13$ and 14, treating $L=15$ as ``exact'' solution. We see that the relative error reaches small values of $10^{-5}$ for $g=50\,\mu$eV and scales as $\propto g^3$ up to $g=0.5$\,meV in agreement with the $g^3$ dependence of the phonon linewidth contribution $\bar{\Gamma}_{\rm ph}$ shown in \Eq{eq:Gamma_FGRph}. Above $\sim 0.5$\,meV, the error saturates at a level below 1\%. Whilst this gives a qualitative picture of the behavior of the error with exciton-cavity coupling strength $g$, one can obtain a more precise estimate of the error by using the exponential dependence on $L$, which is demonstrated for $g=0.6$\,meV in the inset to Fig.\,2(b) of the main text. Deviation from the exponential law and a quicker reduction of the error at larger $L$ seen in the inset is a natural consequence of taking the $L=15$ calculation as exact when evaluating the relative error; if we were to take the true exact solution, we would anticipate a continuation of this exponential trend. One can obviously further refine the estimate of the error by making an extrapolation of all the values of the long-time dependence Eq.\,(26) to $L\to\infty$, using the observed exponential law.

%\begin{figure}[h]
%\hskip-10mm plus 5mm
%\includegraphics[scale=0.35]{Error_L15_calc}
%\caption{Estimated relative error in polariton state linewidths $\Gamma_{1,2}$ at $T$=0 K and $T$=50 K, using the $L$N approach with $L = 15$.}
%\label{fig:Error_L15_calc}
%\end{figure}
%
%Figure~\ref{fig:Error_L15_calc} shows the error in calculation of the linewidths $\Gamma_{1,2}$ via the Trotter decomposition as function of the coupling strength $g$. The error was estimated by using its exponential dependence on $L$, see an example for $g=0.6$\,meV shown in the inset to Fig.\,2(b) of the main text. We see that the relative error reaches small values of $10^{-5}$ for $g=50\,\mu$eV and scales as $\propto g^3$ up to $g=0.5$\,meV where it saturates at a level below 1\%.

%\bibliography{CQED-phonon1.bib}

%\bibliography{CQED-phonon.bib}

%\end{document} 

%\bibliography{CQED-phonon}
%\end{document}

\end{document}